\documentclass{cernyrep}

\newcommand{\der}[2]{\frac{\mathrm{d}#1}{\mathrm{d}#2}}
\newcommand{\pder}[2]{\frac{\partial#1}{\partial#2}}
\newcommand{\rd}[0]{\mathrm{d}}
\usepackage{varwidth}
\usepackage{xcolor}

\begin{document}
\title{Longitudinal Beam Dynamics}
\author{F. Tecker}
\institute{CERN, Geneva, Switzerland}
\maketitle

\begin{abstract}
The course gives a summary of longitudinal beam dynamics for both linear and circular accelerators. After discussing different types of acceleration methods and synchronism conditions, it focuses on the particle motion in synchrotrons.
\end{abstract}

\section{Introduction}
The force $\vec{F}$ on a charged particle with a charge $e$ is given by the Newton--Lorentz force
\begin{equation}
\vec{F}=\frac{\mathrm{d}\vec{p}}{\mathrm{d}t} = e \left( \vec{E} + \vec{v} \times \vec{B} \right).
\label{eq:force}
\end{equation}

The second term is always perpendicular to the direction of motion, so it does not give any longitudinal acceleration and it does not increase the energy of the particle.
Hence, the acceleration has to be done by an electric field $\vec{E}$. In order to accelerate the particle, the field needs to have a component in the direction of the motion of the particle. If we assume the field and the acceleration to be along the $z$ direction, Eq.~(\ref{eq:force}) becomes
\begin{equation}
\der{p}{t} = e E_z.
\end{equation}

The total energy $E$ of a particle is the sum of the rest energy $E_0$ and the kinetic energy $W$:
\begin{equation}
E = E_0 + W.
\end{equation}
In relativistic dynamics, the total energy $E$ and the momentum $p$ are linked by
\begin{equation}
E^2 = E_0^2 + p^2 c^2,
\label{eq:E}
\end{equation}
from which it follows that
\begin{equation}
\mathrm{d}E = v \, \mathrm{d}p.
\end{equation}

The rate of energy gain per unit length of acceleration (along the $z$ direction) is then given by
\begin{equation}
\der{E}{z} = v \der{p}{z} = \der{p}{t} = e E_z
\end{equation}
and the (kinetic) energy gained from the field along the $z$ path follows from $~\mathrm{d}W = \mathrm{d}E = e E_z \,\mathrm{d}z$:
\begin{equation}
W = e \int\! E_z \,\mathrm{d}z = eV,
\end{equation}
where $V$ is just an electric potential.

\section{Methods of acceleration}
\subsection{Electrostatic acceleration}
\begin{figure}[htb]
\centerline{\includegraphics[width=8cm]{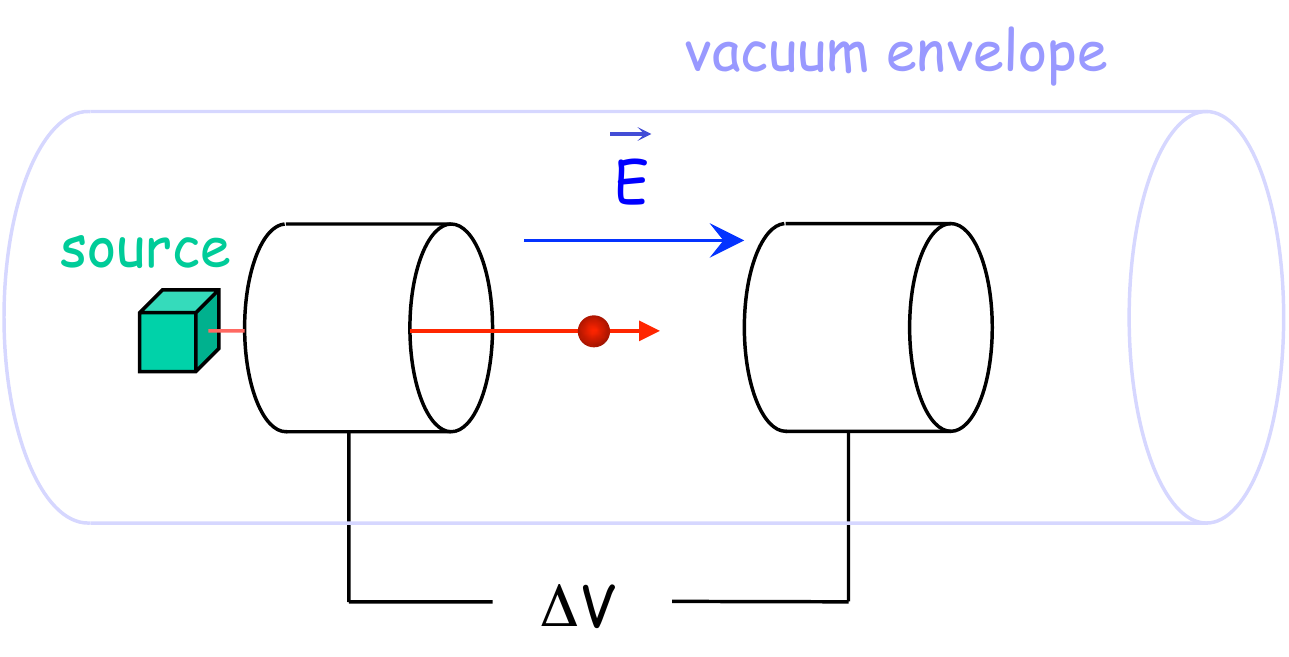}}
 \caption{\label{fig:static} Electrostatic acceleration}
\end{figure}

The most basic way of acceleration is using an electrostatic field between two electrodes, as shown in Fig.~\ref{fig:static}.
The energy gain $W$ in an electrostatic field is given by $W=e \Delta V$, where $\Delta V$ is the voltage (or potential) difference between the electrodes.
This method allows the acceleration of a continuous beam.

On today's energy scale, the maximum energy gain is quite limited by insulation problems, and the maximum voltage is around 10\ MV.
Nevertheless, this is used for the first stage of acceleration, the particle sources, electron guns, X-ray tubes, and low-energy-ion acceleration.

\subsection{Induction -- the betatron}
Insulation issues limit the acceleration by static electric fields. In the general case, the electric field is derived from a scalar potential $V$ and the time derivative of a vector potential $\vec{A}$:
\begin{eqnarray}
\vec{E} & = & - \vec\nabla V - \frac{\partial\vec{A}}{\partial{t}}, \\
\vec{B} &= & \mu \vec{H} = \vec\nabla \times \vec{A} \qquad \Rightarrow \qquad \vec\nabla \times \vec{E} = -  \frac{\partial\vec{B}}{\partial{t}}.
\end{eqnarray}
From Maxwell's equations above, it follows that the time variation of the magnetic field generates an electric field, which can accelerate particles overcoming the static insulation problems.

\begin{figure}[!b]
 \centering\includegraphics[width=0.5\textwidth]{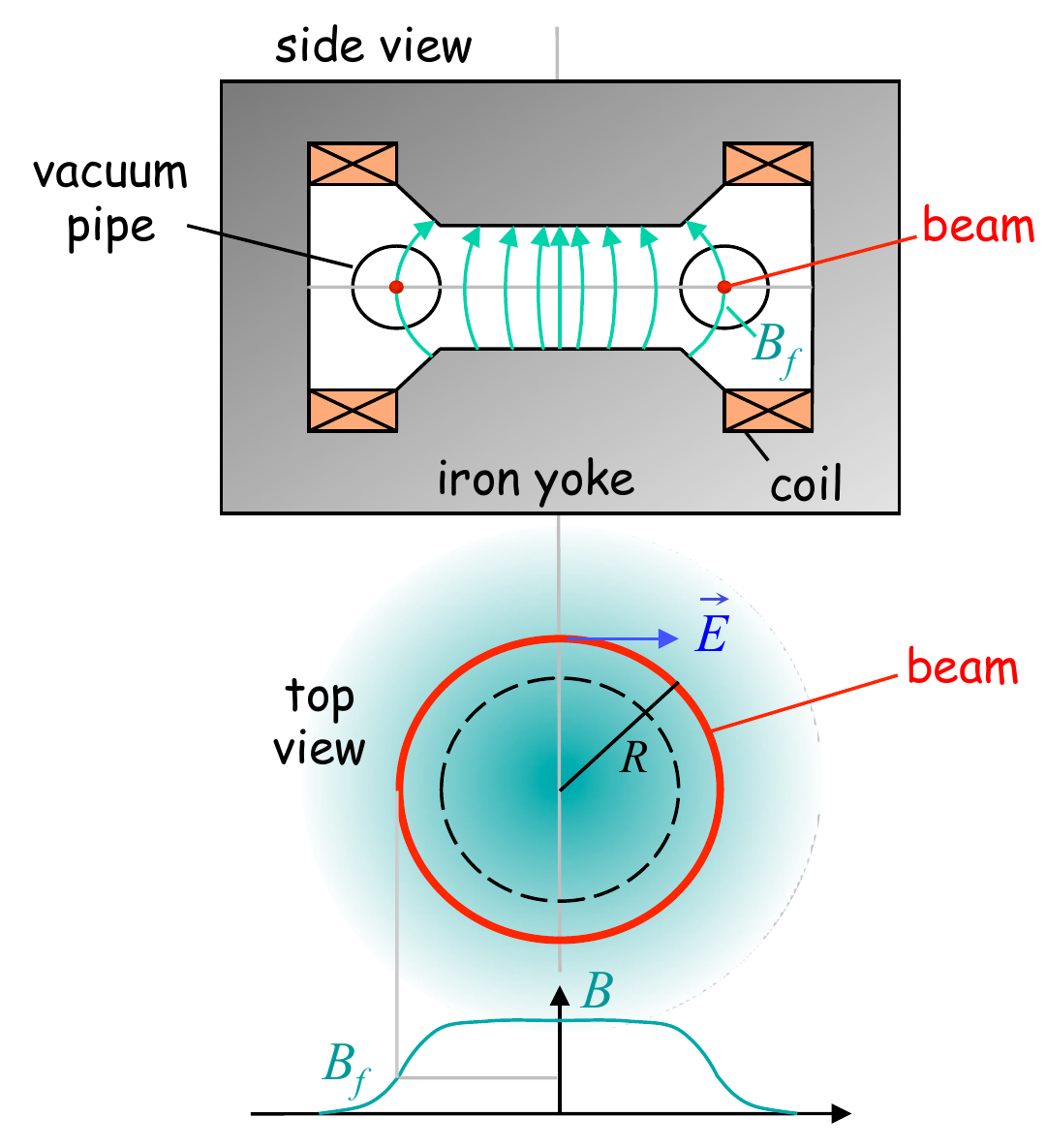}
 \caption{\label{fig:betatron} Schematic of a betatron. The top shows the side view, the middle the top view, and the graph at the bottom the magnetic field distribution.}
\end{figure}

One method based on this principle is the betatron, as shown in Fig.~\ref{fig:betatron}.
The circularly symmetric magnet is fed by an alternating current at a frequency typically between 50 and 200~Hz.
The time-varying magnetic field $\vec{B}$ creates an electric field $\vec{E}$ and at the same time guides the particles on a circular trajectory.
According to the field symmetry, the electric field generated is tangent to the circular orbit.
Betatrons were used to accelerate electrons up to about 300\ MeV with the energy reach limited by the saturation in the magnet yoke.

\subsection{Radio-frequency acceleration}
\begin{figure}[h]
\begin{center}
  \includegraphics[width=0.85\textwidth]{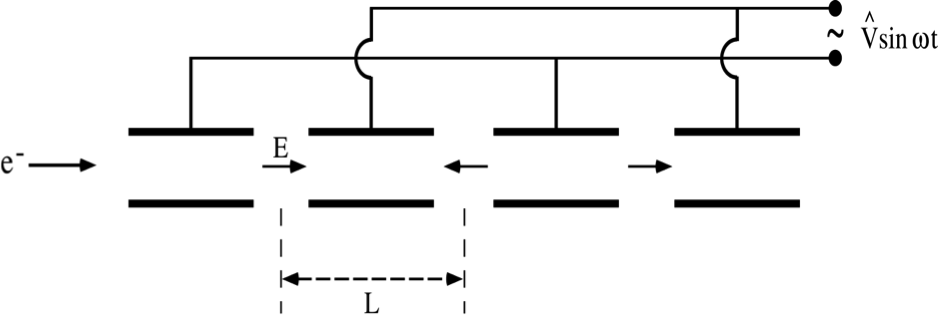}
\caption{Wider\o e-type accelerating structure}
\label{fig:wideroe}
\end{center}
\end{figure}

One also can overcome the limitations of the electrostatic fields by radio-frequency (RF) acceleration.
An RF oscillator feeds alternately a series of drift tubes with gaps in between them, as shown in Fig.~\ref{fig:wideroe}. Inside the tubes, the particle is shielded from the outside field. If the polarity of the field is reversed while the particle travels inside the tube, it gets accelerated at each gap.

This leads to the synchronism condition that the distance $L$ between the gaps has to fulfil:
\begin{equation}
L = v\, T/2,
\end{equation}
where $v = \beta \, c$ is the particle velocity and $T$ the period of the RF oscillator. It is clear that this arrangement cannot accelerate a continuous beam. Only a certain phase range will be accelerated and the beam has to be bunched.

As the particle velocity increases, the drift spaces have to get longer and one loses efficiency. One can increase the radio frequency to counteract this effect but a large amount of power will be radiated as one goes to higher frequencies.
It is then convenient to enclose the accelerating gap in a cavity which holds the electromagnetic energy in the form of a magnetic field
and to make the resonant frequency of the cavity equal to that of the accelerating field.

\begin{figure}[!b]
 \centering\includegraphics[width=0.9\textwidth]{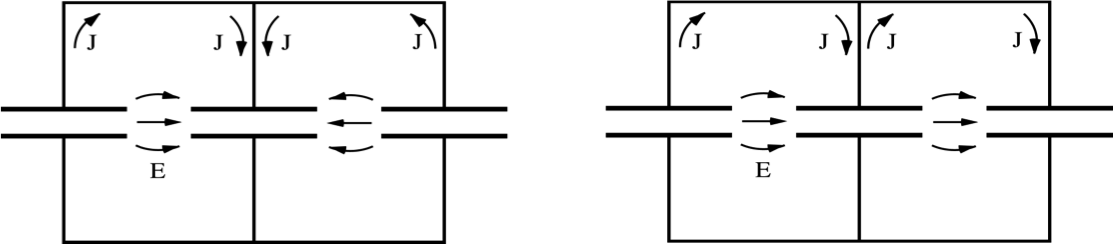}
 \caption{\label{fig:cavities} Adjacent cavities with different modes. Left: $\pi$-mode -- the field is opposite in the gaps of the two cavities, right: $2\pi$-mode -- the field is the same in both cavities.}
 \end{figure}

Several of these cavities can be close together with a certain phase relationship between them (see Fig.~\ref{fig:cavities}).

The synchronism condition depends on the mode. While it is $L = v\, T/2$ for the $\pi$-mode, it becomes $L = v\, T$ for the $2\pi$-mode.
In the $2\pi$-mode, the resulting wall current between the cavities is zero, and the common cavity walls are unnecessary.

\begin{figure}[tb]
 \centering\includegraphics[width=0.95\textwidth]{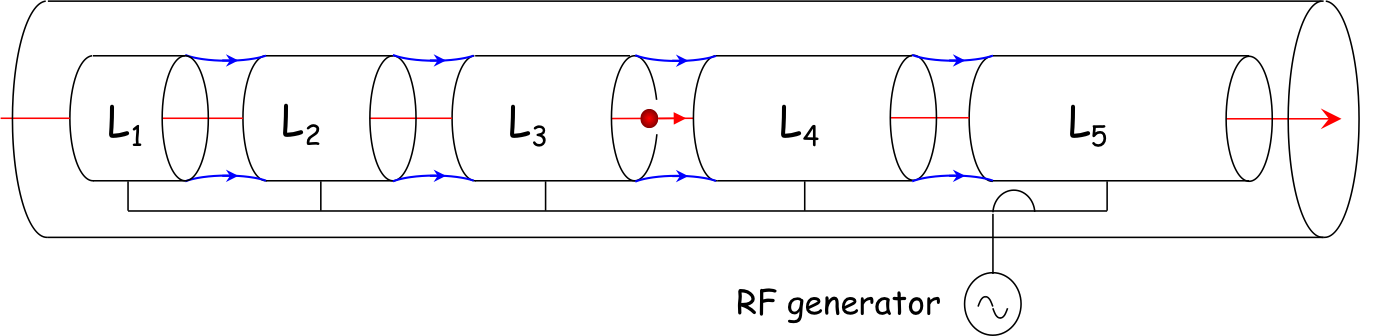}
 \caption{\label{fig:Alvarez} Alvarez-type accelerating structure}
\end{figure}

A variant of that scheme consists of placing the drift tubes in a single resonant tank such that the field has the same phase in all gaps (see Fig.~\ref{fig:Alvarez}). Such a resonant accelerating structure was invented by Alvarez, and this type is still used for the acceleration of protons with the energy ranging from 50 to 200~MeV.

\subsection{Transit-time factor}
When the particle traverses a cavity, the field varies during the passage of the particle through the accelerating gap. So, the particle will not always see the maximum field and the effective acceleration will be smaller by a certain factor.
This {\it transit-time factor} $T_{\rm a}$ is defined as
\begin{equation}
T_{\rm a} = \frac{\mathrm{energy~gain~of~particle~with~}v=\beta c}{\mathrm{maximum~energy~gain~(particle~with~}v \rightarrow \infty)}.
\end{equation}
It quantifies the reduction of energy gain due to the fact that the particle travels with a finite velocity in an electric field with a sinusoidal time variation.
The transit-time factor varies between 0 and 1.

In the general case for a particle travelling in the $z$ direction, assuming that the particle velocity is constant, it is given by
\begin{equation}
T_{\rm a} =  \left|\,\frac{ \displaystyle\int\limits_{-\infty}^{+\infty} E_z(z) \,{\rm e}^{{\rm i}\omega_{\rm RF}t} \,\mathrm{d}z }{\displaystyle\int\limits_{-\infty}^{+\infty} E_z(z) \, \mathrm{d}z} \,\right|.
\end{equation}

\begin{figure}[h]
\hfill 
\begin{minipage}{0.4\textwidth}
  \begin{centering}
    \includegraphics[width=6cm]{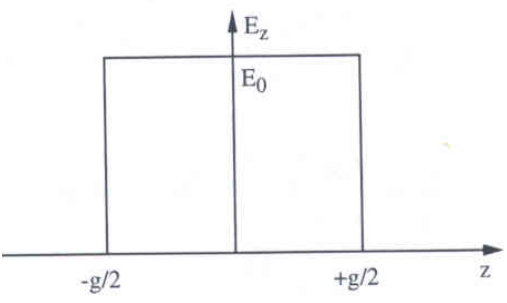}
    \caption{\label{fig:simplefield}Simple uniform field model}
  \end{centering}
\end{minipage}
\begin{minipage}{0.5\textwidth}
  \begin{equation} E_1(z,r) = E_0 = \mathrm{const.} \end{equation} 
\end{minipage}
\hfill
\end{figure}

For a simple model, a uniform standing wave field $E(z,r,t) = E_1(z,r) \cdot \cos(\omega_{\rm RF}t)$ with a constant field only present in the gap (see Fig.~\ref{fig:simplefield}),\\

For a particle passing the centre of the field when the field is maximum, it follows that
\begin{equation}
T_{\rm a} = \left| \, \sin\left(\frac{\omega_{\rm RF} \, g}{2 \, v}\right) \middle/  \, \left(\frac{\omega_{\rm RF} \,g}{2 \, v}\right) \,\right|.
\end{equation}
This simple example shows that the transit-time factor tends towards 1 
for smaller gap width $g$, smaller radio frequency, and higher velocity $v$ of the particle, which is also true in the general case.
The reduction in acceleration becomes important for particles with low velocities, like low-energy protons and particularly ions.

\subsection{Disc-loaded travelling wave structures}
Electrons reach a relativistic $\beta$ close to unity for a kinetic energy of 10~MeV, while protons reach this only at an energy of the order of 10~GeV. Above these energies, the particles have basically the speed of light, and the drift-tube length remains constant. Nevertheless, the drift-tube length would become very long unless the frequency is in the GHz range.

\begin{figure}[htbp]
\begin{center}
\includegraphics[width=0.6\textwidth,height=4.4cm]{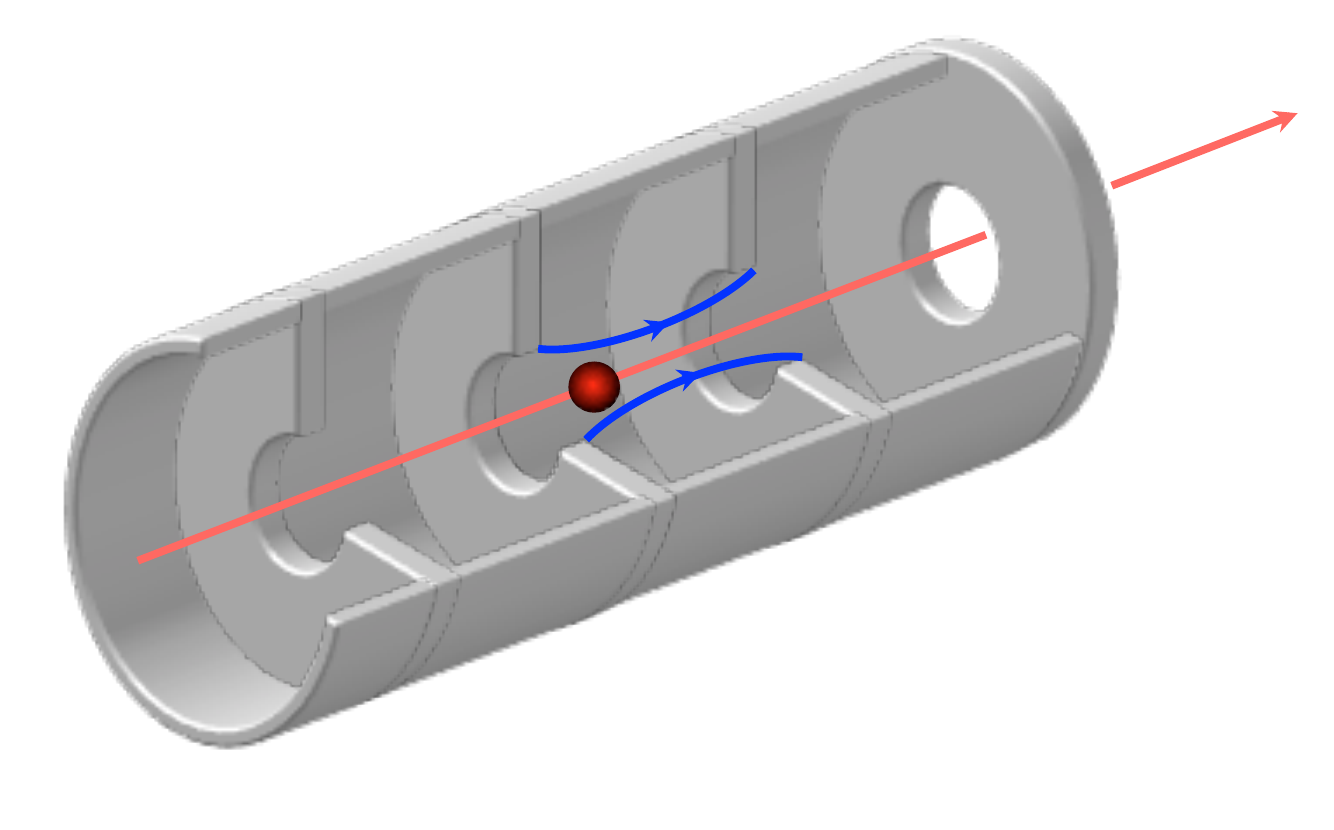}
\caption{Cut of a disc-loaded travelling wave cavity}
\label{fig:tw-cavity}
\end{center}
\end{figure}

So, the idea came up that the ultra-relativistic particles could be accelerated by a travelling wave in a waveguide. In order to to get continuous acceleration, the phase velocity $v_\varphi$ of the wave has to match the velocity $v$ of the particle. However, rectangular or cylindrical waveguides have electric field modes with an electric field in the direction of propagation with phase velocities bigger than $c$, so that the wave does not remain synchronous with the particle.
The phase velocity can be adjusted by inserting irises into the waveguide, and the dimensions of the irises and cells can be tailored to match the phase velocity to the velocity of the particle. Figure~\ref{fig:tw-cavity} shows a sketch of such a disc-loaded travelling wave structure.

\begin{figure}[h] 
\begin{minipage}{0.5\textwidth}
\begin{center}
  \vspace{15pt}\includegraphics[width=\textwidth]{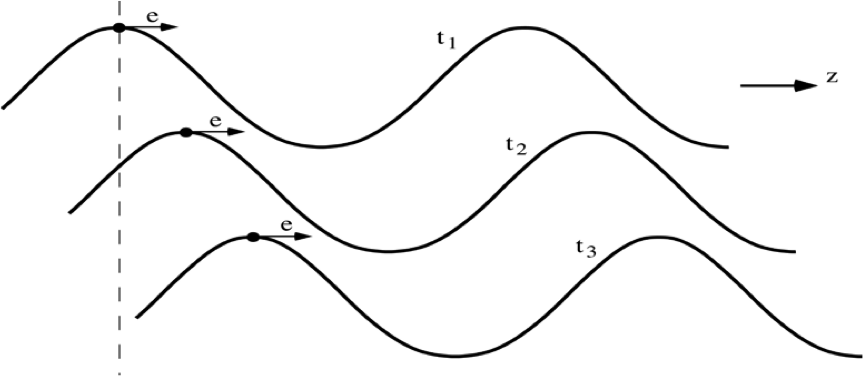}
  \caption{\label{fig:wave} Electromagnetic wave travelling in the $z$ direction}
\end{center}
\end{minipage}
\begin{minipage}{0.5\textwidth}
\begin{eqnarray}
 E_z & = & E_0 \cos(\omega_{\rm RF}t-kz), \label{eq:wave}\\[1mm]
k & = & \frac{\omega_{\rm RF}}{v_\varphi} \qquad \mathrm{wave~number},\nonumber\\
z & = & v \, (t-t_0), \nonumber\\
v_\varphi & = & \mathrm{phase~velocity}, \nonumber\\
v & = & \mathrm{particle~velocity}. \nonumber
\end{eqnarray}
\end{minipage}
\end{figure}

The electric field of an electromagnetic wave travelling in the $z$ direction (see Fig.~\ref{fig:wave}) is given by equation~(\ref{eq:wave}).

The field seen by the particle is
\begin{equation}
E_z = E_0 \cos\!\left( \omega_{\rm RF} t - \omega_{\rm RF} \frac{v}{v_\varphi} t - \phi_0 \right).
\end{equation}

When synchronism is satisfied with $v = v_\varphi$, the particle sees a constant field
\begin{equation}
E_z = E_0 \cos\phi_0,
\end{equation}
where $\phi_0$ is the RF phase seen by the particle. So, this type of structure will continuously accelerate the particle during the passage through the structure.

\section{Phase stability and energy-phase oscillation in a linac}

\begin{figure}[!h]
 \centering\includegraphics[width=0.85\textwidth]{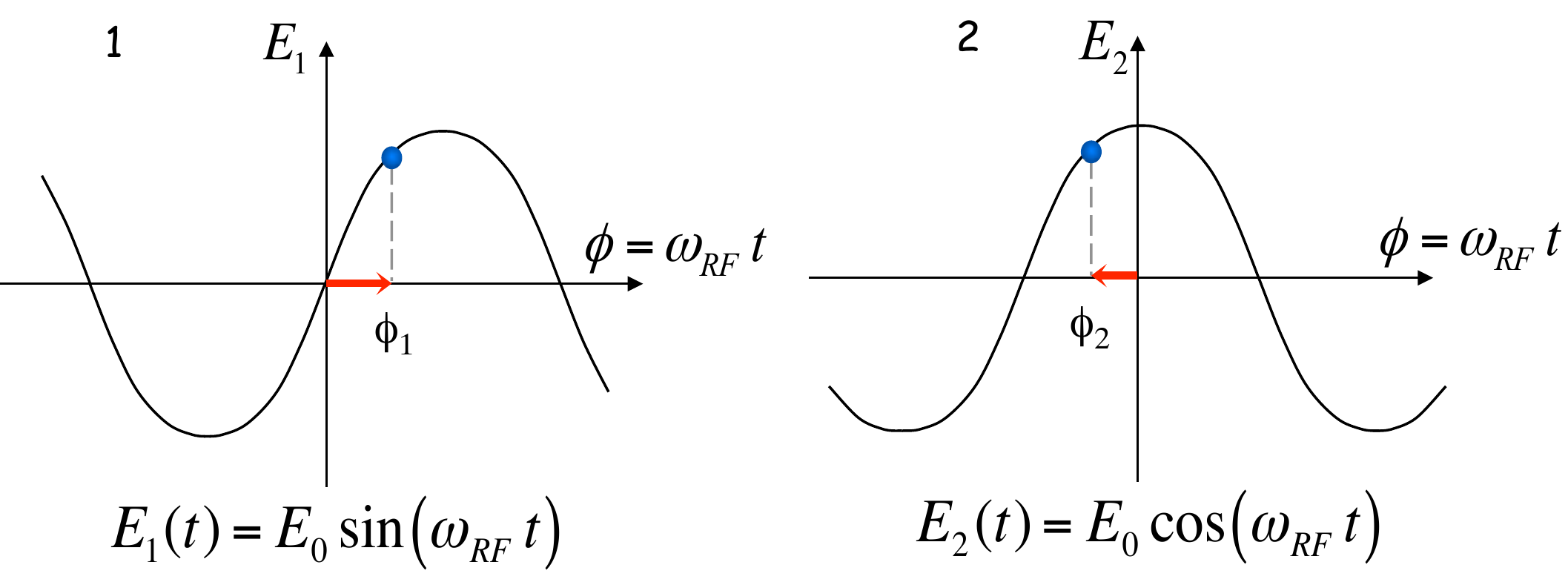}
 \caption{\label{fig:phase-conventions} Common phase conventions: (1) the origin of time is taken at the zero crossing with positive slope, (2) the origin of time is taken at the positive crest of the RF voltage.}
\end{figure}

Several phase conventions exist in the literature (see Fig.~\ref{fig:phase-conventions}):
\begin{itemize}
\item mainly for circular accelerators, the origin of time is taken at the zero crossing with positive slope;
\item mainly for linear accelerators, the origin of time is taken at the positive crest of the RF voltage.
\end{itemize}
In the following, I will stick to the former case of the positive zero crossing for both the linear and circular cases.

\begin{figure}[!tb]
 \centering\includegraphics[width=0.9\textwidth]{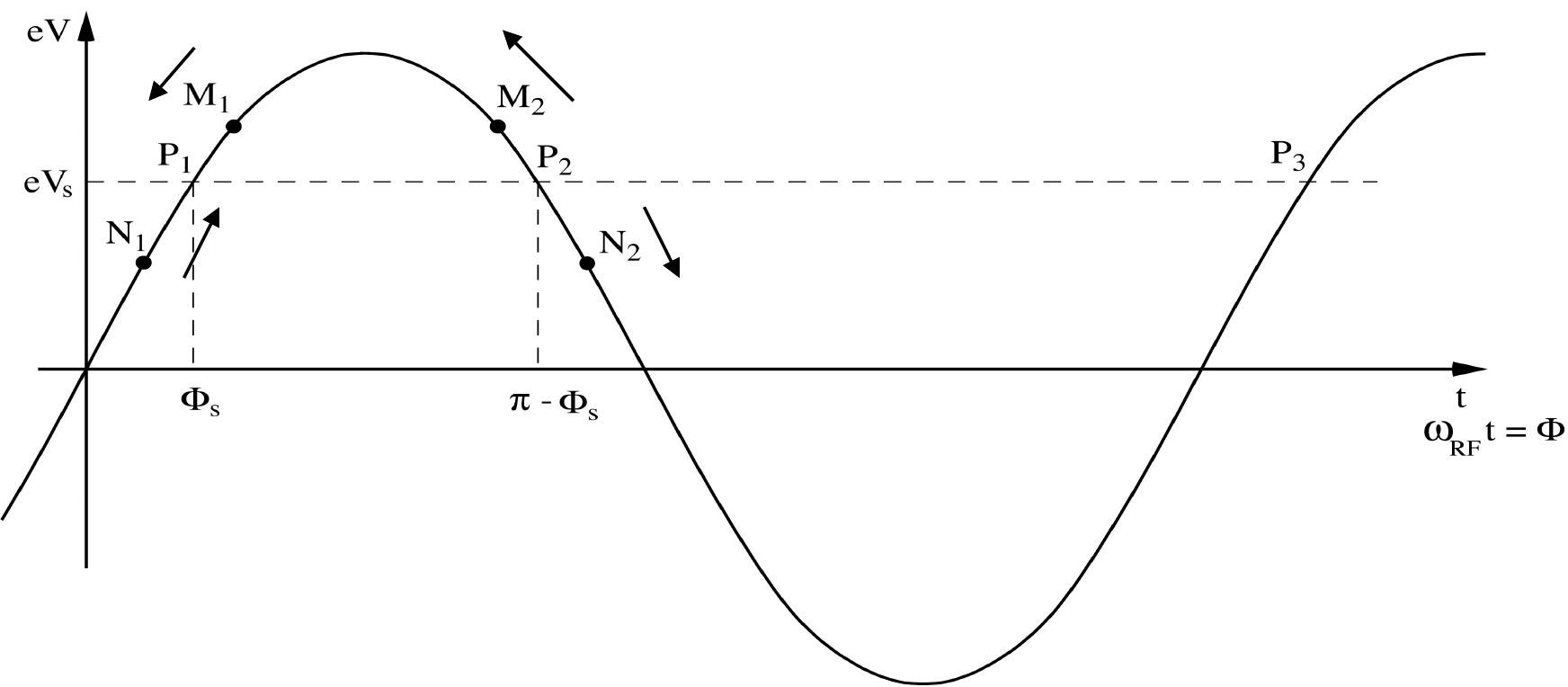}
 \caption{\label{fig:stability} Energy gain as a function of particle phase}
\end{figure}

Let us consider an Alvarez structure, where by design the energy gain for a particle passing through the structure at a certain RF phase $\phi_{\rm s}$ is such that the particle reaches the next gap with the same phase $\phi_{\rm s}$. Then the energy gain in the following gap will be again the same, and the particle will pass all gaps at this phase $\phi_{\rm s}$, which is called the `synchronous phase'. So, the energy gain is $e V_{\rm s} = e \hat{V} \sin \phi_{\rm s}$. This is illustrated in Fig.~\ref{fig:stability} by the points P$_1$ and P$_3$.

A particle N$_1$ which arrives in a gap earlier compared to P$_1$ will gain less energy and
its velocity will be smaller, so that it will take more time to travel through the drift tube. In the
next gap it will appear closer to particle P$_1$. The effect is opposite for particle M$_1$, which will gain
more energy and reduce its delay compared to P$_1$. So, the points P$_1$, P$_3$, etc are stable points
for the acceleration since particles slightly away from them will experience forces that will
reduce their deviation. On the contrary, it can be seen that the point P$_2$ is an unstable point
in the sense that particles slightly away from this point will deviate even more in the next gaps.

So, for stability of the longitudinal oscillation, the particle needs to be on the rising slope of the RF field to have a restoring force towards the stable phase.

\begin{figure}[!hb]
 \centering\includegraphics[width=0.98\textwidth]{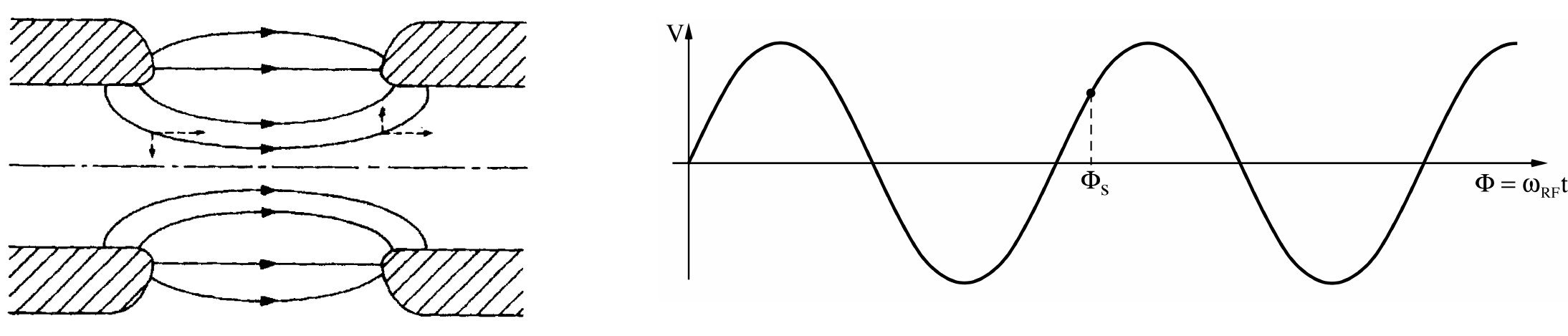}
 \caption{\label{fig:transverse-field} Field lines in the gap of a drift-tube accelerator (left), stable phase on the rising slope of the RF field (right).}
\end{figure}

When we look at the electric field in the accelerating gap between two drift tubes, there is a transverse focusing field at the entrance and a transverse defocusing field at the exit, as illustrated in Fig.~\ref{fig:transverse-field}.
In an electrostatic accelerator, the effect of the defocusing at the exit is smaller than the focusing at the entrance, as the particle gains longitudinal momentum inside the gap.
So, there is a net focusing effect.

In the RF case with stable longitudinal motion, the field increases during the passage of the particle.
As a consequence, the defocusing field when the particle exits the gap is stronger than the focusing field when the particle enters, resulting in a net defocusing effect.
In order to keep the transverse motion stable, external focusing by solenoids or quadrupole magnets is necessary.

In order to study the longitudinal motion it is convenient to use  variables which give the phase
and energy relative to the synchronous particle (denoted by the subscript s):
\begin{eqnarray}
\varphi & = & \phi - \phi_{\rm s}, \\
w & = & E - E_{\rm s} = W - W_{\rm s}.
\end{eqnarray}
The accelerating field can be simply described by
\begin{equation}
E_z = E_0 \sin(\omega t).
\end{equation}

The rate of energy gain for the synchronous particle is given by
\begin{equation}
\frac{\mathrm{d}E_{\rm s}}{\mathrm{d}z} = \der{p_{\rm s}}{t} = e E_0 \sin \phi_{\rm s}
\end{equation}
and for a non-synchronous particle (for small $\varphi$)
\begin{equation}
\der{w}{z} = e E_0 \left[ \sin(\phi_{\rm s} + \varphi) - \sin\phi_{\rm s} \right]  \approx e E_0 \cos\phi_{\rm s} \, \varphi.
\label{eq:w}
\end{equation}

The rate of change of the phase with respect to the synchronous particle is, for small deviations,

\begin{equation}
 \der{\varphi}{z} = \omega_{\rm RF} \left[ \der{t}{z} - \left( \!\der{t}{z}\! \right)_{\!\!{\rm s}} \,\right] =
 \omega_{\rm RF}  \left( \frac{1}{v} - \frac{1}{v_{\rm s}} \right) \approx - \frac{\omega_{\rm RF}}{v^2_{\rm s}} (v-v_{\rm s}).
\end{equation}
Using $\rd \gamma = \gamma^3 \beta \, \rd\beta$, $w$ becomes in the vicinity of the synchronous particle

\begin{equation}
w = E -E_{\rm s} = m_0 c^2 ( \gamma - \gamma_{\rm s} ) = m_0 c^2 \,\rd\gamma = m_0 c^2 \gamma_{\rm s}^3 \beta_{\rm s} \, \rd\beta = m_0 \gamma_{\rm s}^3 v_{\rm s} (v - v_{\rm s}),
\end{equation}
which leads to

\begin{equation}
 \der{\varphi}{z} = - \frac{\omega_{\rm RF}}{m_0 v_{\rm s}^3 \gamma_{\rm s}^3} w.
 \label{eq:phi}
\end{equation}

Combining the two first-order equations~(\ref{eq:w}) and (\ref{eq:phi}) into a second-order equation gives the equation of an harmonic oscillator with the angular frequency $\Omega_{\rm s}$:
\begin{equation}
\der{^2\varphi}{z^2} + \Omega_{\rm s}^2 \varphi = 0  \qquad\mathrm{with}\qquad \Omega_{\rm s}^2 = \frac{eE_0\omega_{\rm RF}\cos\phi_{\rm s}}{m_0v_{\rm s}^3\gamma_{\rm s}^3}.
\label{eq:lin-osc}
\end{equation}
Stable harmonic oscillations imply that $\Omega_{\rm s}^2 > 0$ and real, which means that $\cos\phi_{\rm s} > 0$. Since acceleration means that $\sin\phi_{\rm s} >0$, it follows that the stable phase region for acceleration in the linac is

\begin{equation}
0 < \phi_{\rm s} < \frac{\pi}{2},
\end{equation}
which confirms what we have seen before in our discussion about the restoring force towards the stable phase.

From Eq.~(\ref{eq:lin-osc}) it is also visible that the oscillation frequency decreases strongly with growing velocity (and relativistic gamma) of the particle. For highly relativistic particles, the velocity change is negligible, so there is practically no change of the particle phase, and the bunch distribution is not changing any more.

\section{Circular accelerators -- cyclotron}

\begin{figure}[h]
 \centering{\includegraphics[width=0.45\textwidth]{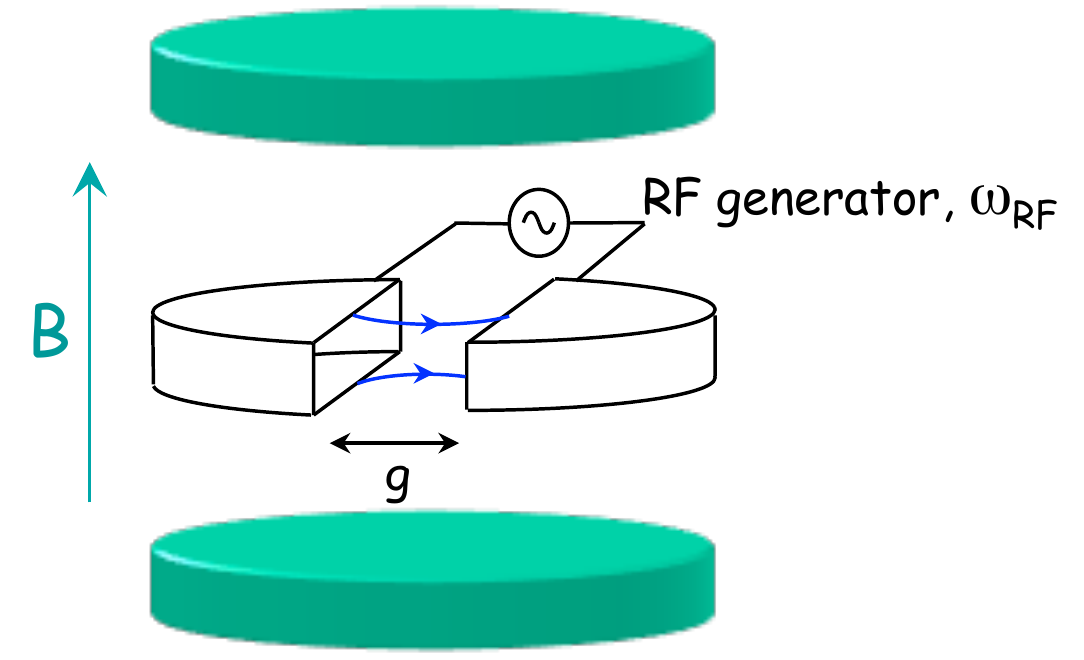}\hskip1cm
\includegraphics[width=0.45\textwidth]{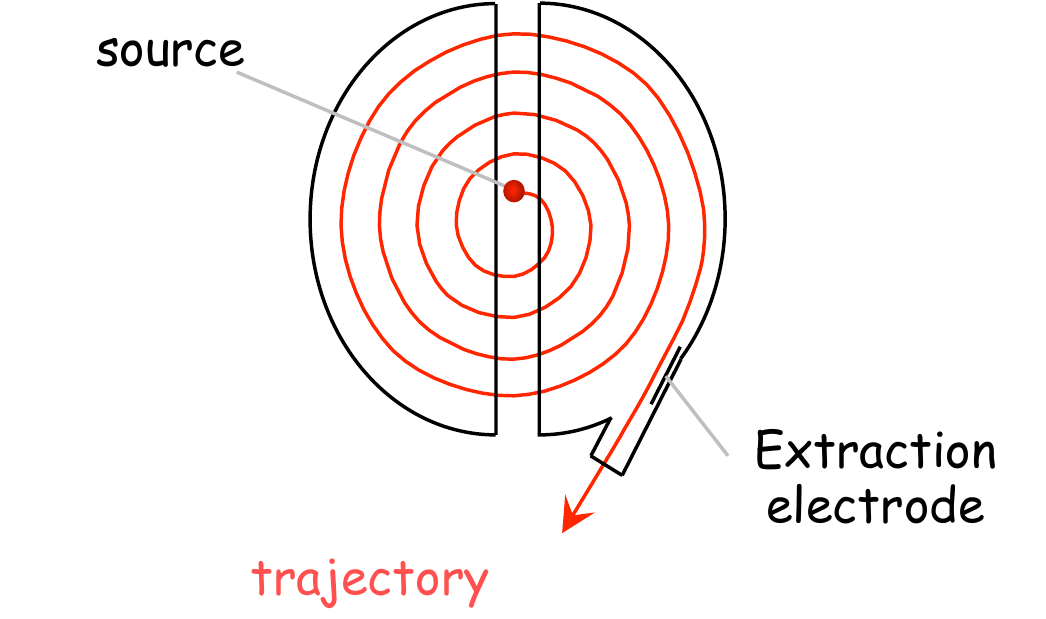}}
 \caption{\label{fig:cyclotron} Schematic view of a cyclotron and the particle trajectory in a top view (on the right)}
\end{figure}

The cyclotron is a circular accelerator that has two hollow `D'-shape electrodes in a constant magnetic field $B$ (see Fig.~\ref{fig:cyclotron}).
When a particle is generated at the source in the centre, it is accelerated by the electric field between the electrodes. It enters an electrode and, while it is shielded from the electric field, the polarity of the field in the gap is reversed. If the phase of the RF is correct, the particle is accelerated again in the gap and enters the other electrode. The magnetic field creates a spiralling trajectory of the particle. As the particle becomes faster, the orbit radius gets bigger but the revolution frequency does not depend on the radius, as long as the particle is not relativistic.

So, the synchronism condition is that the RF period has to correspond to the revolution period:
\begin{equation}
T_{\rm RF} = 2\pi \rho / v_{\rm s}
\end{equation}
with the cyclotron frequency given by
\begin{equation}
\omega = \frac{q B}{m_0 \gamma}.
\end{equation}
As long as $v\ll c$ and $\gamma \approx 1$ the synchronism condition stays fulfilled. For higher energies, the particle will get out of phase with respect to the RF, even though there is still a range for the initial particle phase where a stable acceleration is possible.

In order to keep synchronism at higher energies, one has to decrease the radio frequency during the acceleration cycle according to the relativistic $\gamma(t)$ of the particle as
\begin{equation}
\omega_{\rm RF}(t) = \omega_{\rm s}(t) = \frac{q B}{m_0 \gamma(t)},
\end{equation}
which leads to the concept of a synchrocyclotron, which can accelerate protons up to around 500~MeV. Here a new limitation occurs due to the size of the magnet.

\section{Synchrotron}
\begin{figure}[h]
 \centering\includegraphics[width=0.5\textwidth]{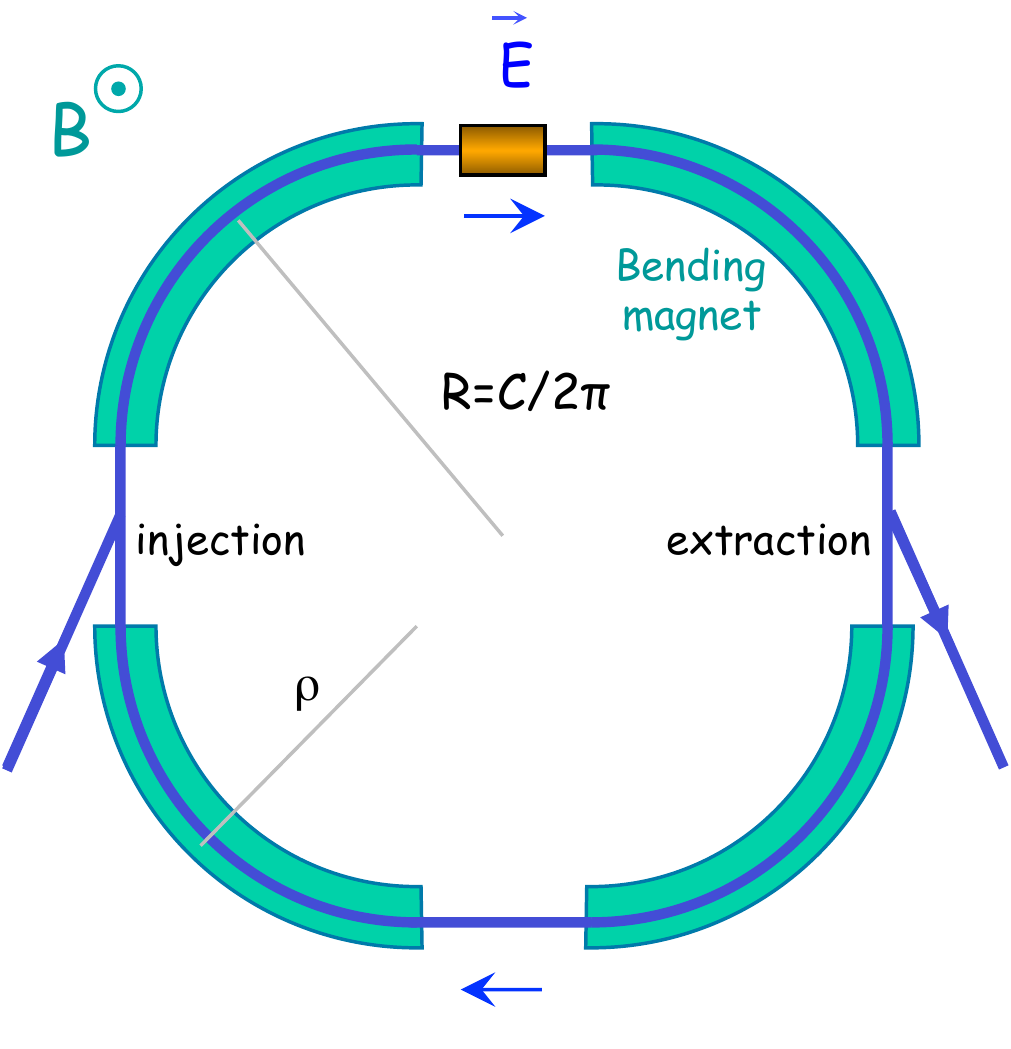}
 \caption{\label{fig:synchrotron} Schematic layout of a synchrotron}
\end{figure}

A synchrotron (see Fig.~\ref{fig:synchrotron}) is a circular accelerator where the nominal particle trajectory is kept at a
constant physical radius by variation of both the magnetic field and the radio frequency, in order to follow the energy variation.
In this way, the aperture of the vacuum chamber and the magnets can be kept small.

The radio frequency needs to be synchronous to the revolution frequency. To achieve synchronism, the synchronous particle needs to arrive at the cavity again after one turn with the same phase. This implies that the angular radio frequency $\omega_{\rm RF} = 2\pi f_{\rm RF}$ has to be an integer multiple of the angular revolution frequency:
\begin{equation}
\omega_{\rm RF} = h \, \omega_{\rm r},
\end{equation}
where $h$ is an integer and is called the {\it harmonic number}. As a consequence, the number of stable synchronous particle locations equals the harmonic number $h$. They are equidistantly spaced around the circumference of the accelerator. All synchronous particles will have the same nominal energy and will follow the nominal trajectory.

Energy ramping is obtained by varying the magnetic field, while following the change of the revolution frequency with a change of the radio frequency.
The time derivative of the momentum,
\begin{equation}
p = e B \rho,
\end{equation}
yields (when keeping the radius $\rho$ constant)
\begin{equation}
\der{p}{t} = e \rho \dot{B}.
\end{equation}
For one turn in the synchrotron, this results in
\begin{equation}
(\Delta p)_\mathrm{turn} = e \rho \dot{B} T_{\rm r} = \frac{2 \pi e \rho R \dot{B}}{v},
\end{equation}
where $R=L / 2\pi$ is the physical radius of the machine.

Since $E^2 = E_0^2 + p^2 c^2$,
it follows that
$\Delta E = v \Delta p$, so that
\begin{equation}
(\Delta E)_\mathrm{turn} = (\Delta W)_{\rm s} = 2 \pi e \rho R \dot{B} = e \hat{V} \sin \phi_{\rm s}.
\end{equation}

From this relation it can be seen that the stable phase for the synchronous particle changes during the acceleration, when the magnetic field $B$ changes, as
\begin{equation}
\sin \phi_{\rm s} = 2 \pi \rho R \frac{\dot{B}}{\hat{V}_{\rm RF}} \qquad \mathrm{or} \qquad \phi_{\rm s} = \arcsin\left( 2 \pi \rho R \frac{\dot{B}}{\hat{V}_{\rm RF}} \right).
\end{equation}

As mentioned previously, the radio frequency has to follow the change of revolution frequency and will increase during acceleration as
\begin{equation}
\frac{f_{\rm RF}}{h} = f_{\rm r} = \frac{v(t)}{2\pi R_{\rm s}} = \frac{1}{2\pi} \frac{ec^2}{E_{\rm s}(t)} \frac{\rho}{R_{\rm s}} B(t).
\end{equation}

Since $E^2 = E_0^2 + p^2 c^2$,
the radio frequency must follow the variation of the $B$ field with the law
\begin{equation}
\frac{f_{\rm RF}}{h}  = \frac{c}{2\pi R_{\rm s}} \left\{ \frac{B(t)^2}{(m_0c^2 / e c \rho)^2 + B(t)^2} \right\}^{1/2}.
\end{equation}
This asymptotically tends towards $f_{\rm r} \rightarrow {c}/({2\pi R_{\rm s}})$ when $v \rightarrow c$ and $B$ becomes large compared to $m_0c^2 / (e c \rho)$.

\subsection{Dispersion effects in a synchrotron}
If a particle is slightly shifted in momentum, it will have a different velocity and also a different orbit and orbit length.
We can define two parameters:
\begin{itemize}
\item The {\it momentum compaction factor} $\alpha$, which is the relative change in orbit length with momentum:
\begin{equation}
\alpha = \frac{\Delta L/L}{\Delta p / p}.
\end{equation}
\item The {\it slip factor} $\eta$, which is the relative change in revolution frequency with momentum:
\begin{equation}
\eta = \frac{\Delta f_{\rm r}/f_{\rm r}}{\Delta p / p}.
\end{equation}
\end{itemize}

\begin{tabular}{p{0.4\textwidth} p{0.4\textwidth}}
  \vspace{0pt}\hspace{1cm} \includegraphics[width=0.3\textwidth]{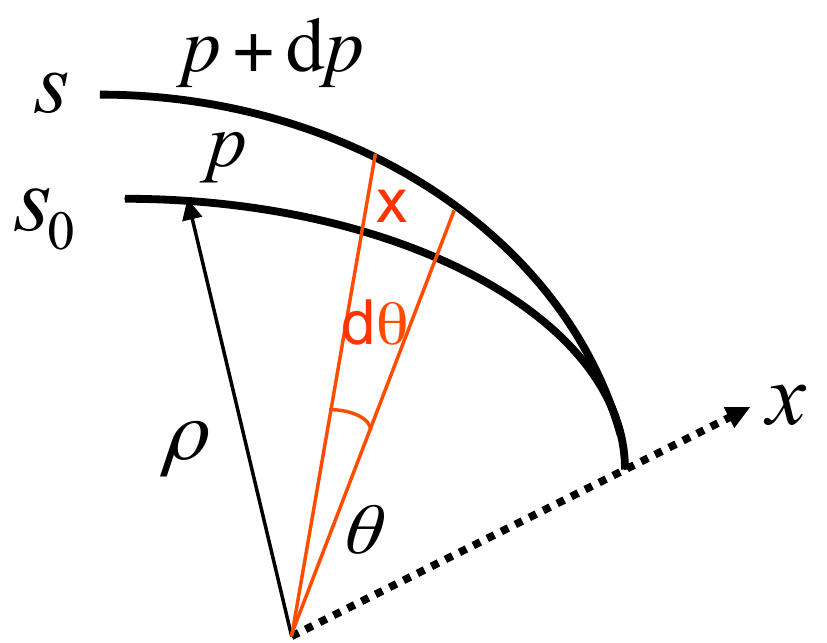} &
 \vspace{20pt}\begin{eqnarray}
\mathrm{d}s_0 & = & \rho \, \mathrm{d}\theta, \nonumber\\
\mathrm{d}s & = & ( \rho + x ) \, \mathrm{d}\theta. \nonumber
\end{eqnarray}
\end{tabular}

Let us consider the change in orbit length. The relative elementary path length difference for a particle with a momentum $p + \mathrm{d}p$ is
\begin{equation}
\der{l}{s_0} = \der{s - \mathrm{d}s_0}{s_0} = \frac{x}{\rho} = \frac{D_x}{\rho} \frac{\mathrm{d}p}{p},
\end{equation}
where $D_x = \mathrm{d}x / (\mathrm{d}p/p)$ is the {\it dispersion function} from the transverse beam optics.

This leads to a total change in the circumference $L$ of
\begin{equation}
\mathrm{d}L = \int_C \mathrm{d}l = \int \frac{x}{\rho} \,\mathrm{d}s_0 = \int \frac{D_x}{\rho} \frac{\mathrm{d}p}{p} \,\mathrm{d}s_0,
\end{equation}
so that
\begin{equation}
\alpha = \frac{1}{L} \int \frac{D_x}{\rho} \,\mathrm{d}s_0.
\end{equation}

Since $\rho = \infty$ in the straight sections, we get
\begin{equation}
\alpha = \frac{<D_x>_{\rm m}}{R},
\end{equation}
where the average $< \quad>_{\rm m}$ is considered over the bending magnets only.

Given that the revolution frequency is $f_{\rm r} = \beta c / 2\pi R$, the relative change is (using the definition of the momentum compaction factor)
\begin{equation}
\frac{\rd f_{\rm r}}{f_{\rm r}} = \frac{\rd\beta}{\beta} - \frac{\rd R}{R} = \frac{\rd\beta}{\beta} - \alpha \frac{\rd p}{p},
\end{equation}

\begin{equation}
p = mv = \beta \gamma \frac{E_0}{c} \Rightarrow \frac{\rd p}{p} =  \frac{\rd\beta}{\beta} + \frac{\rd(1-\beta^2)^{-1/2}}{(1-\beta^2)^{-1/2}} = \underbrace{\left( 1-\beta^2 \right)^{-1}}_{\displaystyle\gamma^2} \frac{\rd\beta}{\beta}.
\end{equation}

So, the relative change in revolution frequency is given by
\begin{equation}
\frac{\rd f_{\rm r}}{f_{\rm r}} = \left( \frac{1}{\gamma^2} - \alpha \right) \frac{\rd p}{p},
\end{equation}
which means that the slip factor $\eta$ is given by

\begin{equation}
\eta = \frac{1}{\gamma^2} - \alpha.
\end{equation}

Obviously, there is one energy with a given $\gamma_{\rm tr}$ for which $\eta$ becomes zero, meaning that there is no change of the revolution frequency for particles with a small momentum deviation. This energy is a property of the transverse lattice with
\begin{equation}
\gamma_{\rm tr} = \frac{1}{\sqrt{\alpha}}.
\end{equation}

From the definition of $\eta$, it is clear that an increase in momentum gives
\begin{itemize}
\item {\it below transition} energy $(\eta > 0)$: a higher revolution frequency. The increase of the velocity of the particle is the dominating effect;
\item {\it above transition} energy $(\eta < 0)$: a lower revolution frequency. The particle has a velocity close to the speed of light and this velocity does not change significantly any more. So, here the effect of the longer path length dominates (for the most common case of transverse lattices with a positive momentum compaction factor,  $\alpha>0$).
\end{itemize}

At transition, the velocity change and the path-length change with momentum compensate each other.
So, the revolution frequency there is independent of the momentum deviation.
As a consequence, the longitudinal oscillation stops and the particles in the bunch will not change their phase.
Particles which are not at the synchronous phase will get the same non-nominal energy gain each turn and will accumulate an energy error that will increase the longitudinal emittance and can lead to a loss of the particle due to dispersive effects. So, transition has to be passed quickly in order to minimise the emittance increase and the losses.

A method to improve transition crossing is to change the transverse optics when the energy almost reaches $\gamma_{\rm tr}$ for an optics with a bigger momentum compaction factor $\alpha$. Hence, $\gamma_{\rm tr}$ is decreasing at the same time as the energy is increasing, and the time with an energy close to transition can be reduced.

Electron synchrotrons do not need to cross transition. Due to the relatively low rest mass of the electron, the relativistic gamma is so high that already the injection energy is above the transition energy. Hence, the electrons will stay above transition during the whole acceleration cycle.

\begin{figure}[h]
 \centering\includegraphics[width=\textwidth]{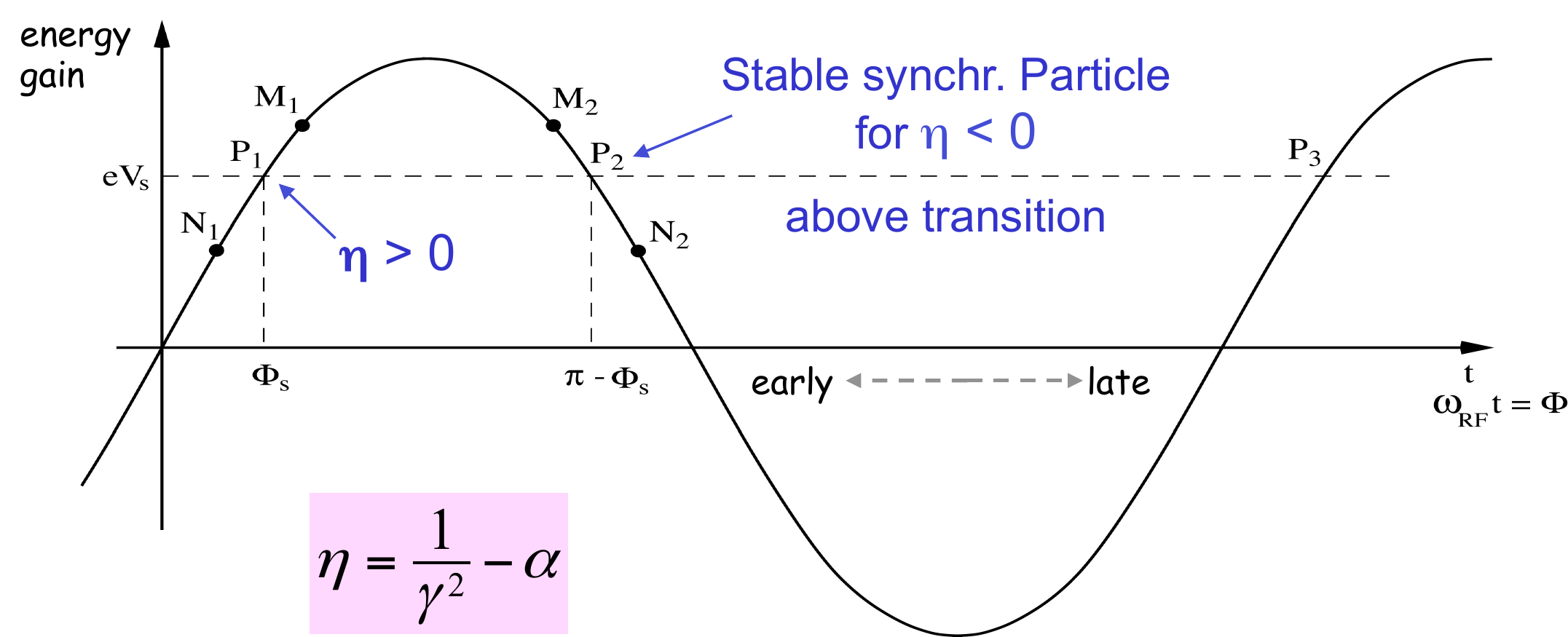}
 \caption{\label{fig:phase-stability} Energy gain as a function of particle phase. The oscillations are stable around the synchronous phase particle P$_1$ below transition and around the synchronous phase particle P$_2$ above transition.}
\end{figure}

Since the change of revolution frequency with momentum is opposite below and above transition, this changes completely the range for stable oscillations. As we have seen in the linac case (see Fig.~\ref{fig:stability}), the oscillation is stable for a particle that is on the rising slope of the RF field when we are below transition. Above transition, the oscillation there is unstable and the stable region for oscillations is on the falling slope (see Fig.~\ref{fig:phase-stability}). A particle that arrives too early (M$_2$) will get more energy, and the revolution time will increase due to the predominant effect of the longer path. So, it will arrive later on the next turn, closer to the synchronous phase. Similarly, a particle that arrives late (N$_2$) will gain less energy and travel a shorter orbit, moving as well towards the synchronous phase.

\begin{figure}[h]
 \centering\includegraphics[width=0.6\textwidth]{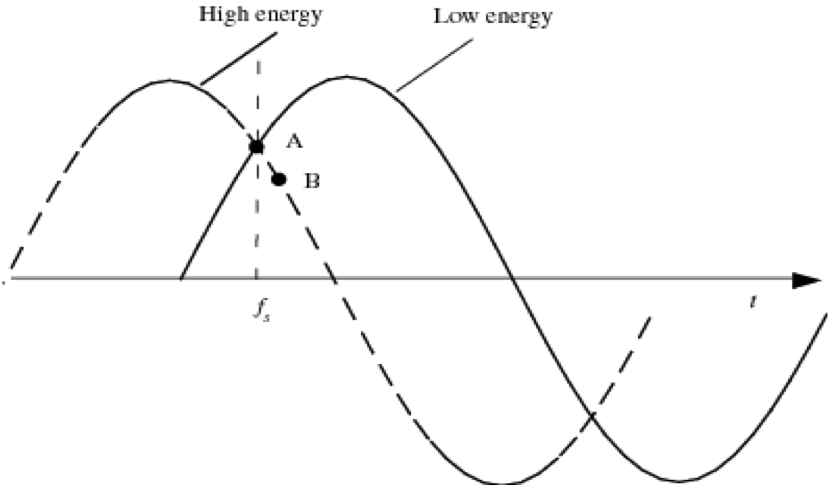}
 \caption{\label{fig:transition} The synchronous phase changes from rising to falling slopes as the energy crosses transition \cite{wilson}}
\end{figure}

Crossing transition during acceleration makes the previous stable synchronous phase unstable. Below transition, it is stable on the rising slope of the RF; above transition, the synchronous phase is on the falling slope. Consequently, the RF system needs to make a rapid change of the RF phase when crossing transition energy, a `phase jump', as indicated in Fig.~\ref{fig:transition}. Otherwise the particles in the bunch get dispersed, have a wrong energy gain, and get eventually lost.

As previously done for the linac, we want to look at the oscillations with respect to the synchronous particle and we express the variables with respect to the synchronous particle:

\begin{tabular}{p{0.5\textwidth} p{0.25\textwidth}}
\vspace{30pt}\begin{tabular}{ll}
particle RF phase: & $\Delta\phi = \phi - \phi_{\rm s}$, \\
particle momentum: & $\Delta p = p - p_{\rm s}$, \\
particle energy: & $\Delta E = E - E_{\rm s}$, \\
azimuth angle:& $\Delta\theta = \theta - \theta_{\rm s}.$\\
\end{tabular}
&
\vspace{0pt}\includegraphics[width=0.25\textwidth]{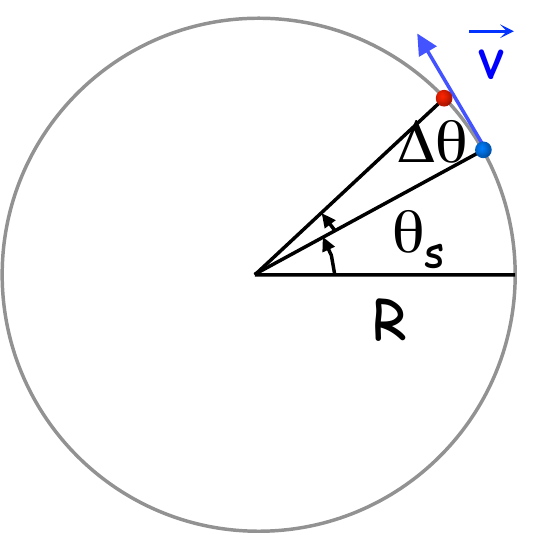}
\end{tabular}

Since the radio frequency is a multiple of the revolution frequency, the RF phase $\Delta\phi$ changes as
\begin{equation}
f_\mathit{\rm RF} = h\, f_{\rm r} \qquad \Rightarrow \qquad \Delta\phi = -h \, \Delta\theta \qquad \mathrm{with} \qquad \theta = \int \omega_{\rm r} \rd t.
\end{equation}
The minus sign for the RF phases originates from the fact that a particle that is ahead arrives earlier, so at a smaller RF phase.

For a given particle with respect to the reference one, the change in angular revolution frequency is
\begin{equation}
\Delta\omega_{\rm r} = \der{}{t} \left( \Delta\theta \right) = - \frac{1}{h} \der{}{t} \left( \Delta\phi \right) = - \frac{1}{h} \der{\phi}{t}.
\end{equation}

Since $\eta = \frac{p_{\rm s}}{\omega_{\rm rs}} \left( \der{\omega_{\rm r}}{p} \right)_{\!{\rm s}}$, $E^2 = E_0^2 + p^2 c^2$, and $\Delta E = v_{\rm s} \Delta p = \omega_{\rm rs} R_{\rm s} \Delta p$, one gets the first-order equation
\begin{equation}
\frac{\Delta E}{\omega_{\rm rs}} = - \frac{p_{\rm s} R_{\rm s}}{h \eta \omega_{\rm rs}} \der{(\Delta\phi)}{t} = - \frac{p_{\rm s} R_{\rm s}}{h \eta \omega_{\rm rs}} \dot{\phi}. \label{eq:first1}
\end{equation}
The second first-order equation follows from the energy gain of a particle:
\begin{equation}
\der{E}{t} = \frac{\omega_{\rm r}}{2\pi} \,e \hat{V} \sin\phi,
\end{equation}
\begin{equation}
2 \pi \der{}{t} \left( \frac{\Delta E}{\omega_{\rm rs}} \right) = e \hat{V} \left( \sin\phi - \sin\phi_{\rm s} \right). \label{eq:first2}
\end{equation}

Deriving and combining the two first-order equations (\ref{eq:first1}) and (\ref{eq:first2}) leads to
\begin{equation}
\der{}{t} \left[ \frac{R_{\rm s} p_{\rm s}}{h \eta \omega_{\rm rs}} \der{\phi}{t} \right] + \frac{e \hat{V}}{2\pi}  \left( \sin\phi - \sin\phi_{\rm s} \right) = 0.
\label{eq:gen2}
\end{equation}
This second-order equation is non-linear and the parameters within the bracket are in general slowly varying in time.

When we assume constant parameters $R_{\rm s}, p_{\rm s}, \omega_{\rm s}$, and $\eta$, we get

\begin{equation}
\ddot\phi + \frac{\Omega_{\rm s}^2}{\cos\phi_{\rm s}}  \left( \sin\phi - \sin\phi_{\rm s} \right) = 0 \qquad \mathrm{with} \qquad \Omega_{\rm s}^2 = \frac{h \eta \omega_{\rm rs} e \hat{V} \cos\phi_{\rm s}}{2\pi R_{\rm s} p_{\rm s}} \label{eq:osc}
\end{equation}
and, for small phase deviations from the the synchronous particle,
\begin{equation}
\sin\phi - \sin\phi_{\rm s} = \sin(\phi_{\rm s}+\Delta\phi) - \sin\phi_{\rm s} \approxeq \cos\phi_{\rm s} \Delta\phi,
\end{equation}
so that we end up with the equation of an harmonic oscillator:
\begin{equation}
\ddot\phi + \Omega_{\rm s}^2 \Delta\phi = 0,
\end{equation}
where $\Omega_{\rm s}$ is the synchrotron angular frequency.

\begin{figure}[h]
 \centering\includegraphics[width=0.85\textwidth]{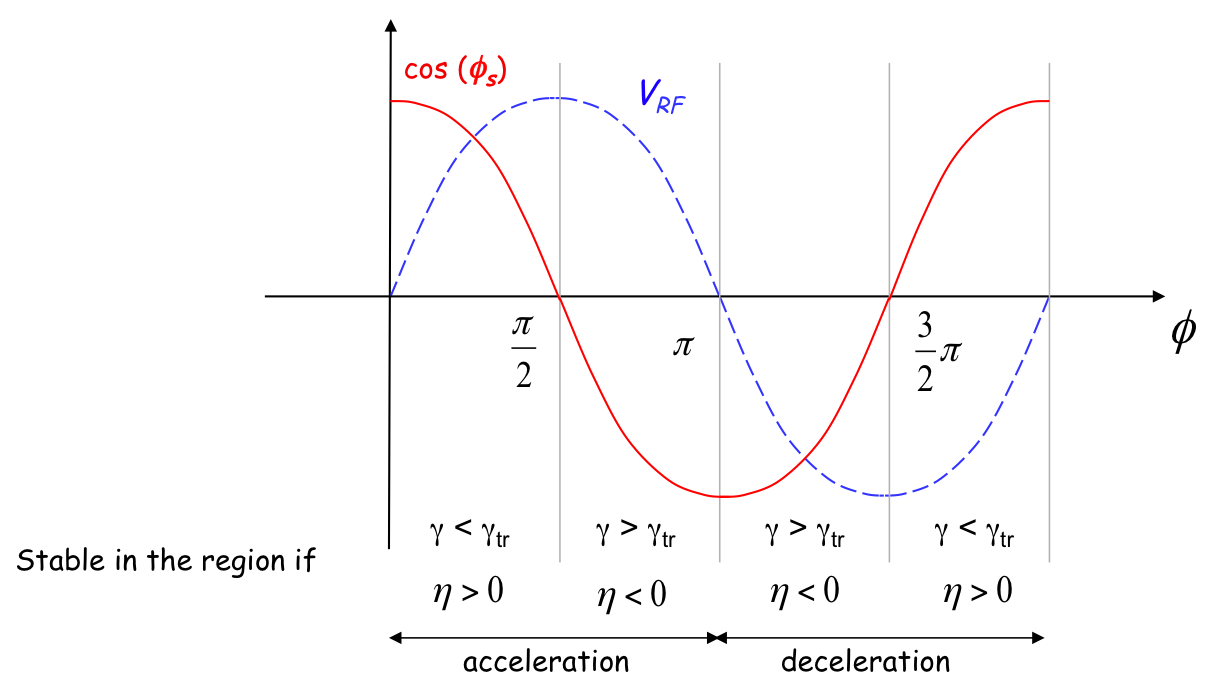}
 \caption{\label{fig:stability-regions} Stability regions as a function of particle phase depending on the energy with respect to transition}
\end{figure}

Stability is obtained when $\Omega_{\rm s}$ is real so that $\Omega_{\rm s}^2$ is positive. Since most terms in $\Omega_{\rm s}^2$ are positive, this reduces to
\begin{equation}
\eta \cos\phi_{\rm s} > 0
\end{equation}
and the stable region for the synchronous phase depends on the energy with respect to the transition energy, as we have seen from our argument before. The conditions for stability are summarised in Fig.~\ref{fig:stability-regions}.

For larger phase (or energy) deviations from the synchronous particle, we can multiply Eq.~(\ref{eq:osc}) by $\dot\phi$ and integrate it, getting an invariant of motion:
\begin{equation}
\frac{\dot\phi^2}{2} - \frac{\Omega_{\rm s}^2}{\cos\phi_{\rm s}}  \left( \cos\phi + \phi \sin\phi_{\rm s} \right) = I,
\label{eq:invariant}
\end{equation}
which, for small amplitudes for $\Delta\phi$, reduces to
\begin{equation}
\frac{\dot\phi^2}{2} + \Omega_{\rm s}^2 \,\frac{(\Delta\phi)^2}{2} = I'.
\end{equation}
Similar equations exist for the second variable $\Delta E \propto \der{\phi}{t}$.

\begin{figure}[htb]
 \centering\includegraphics[width=0.6\textwidth]{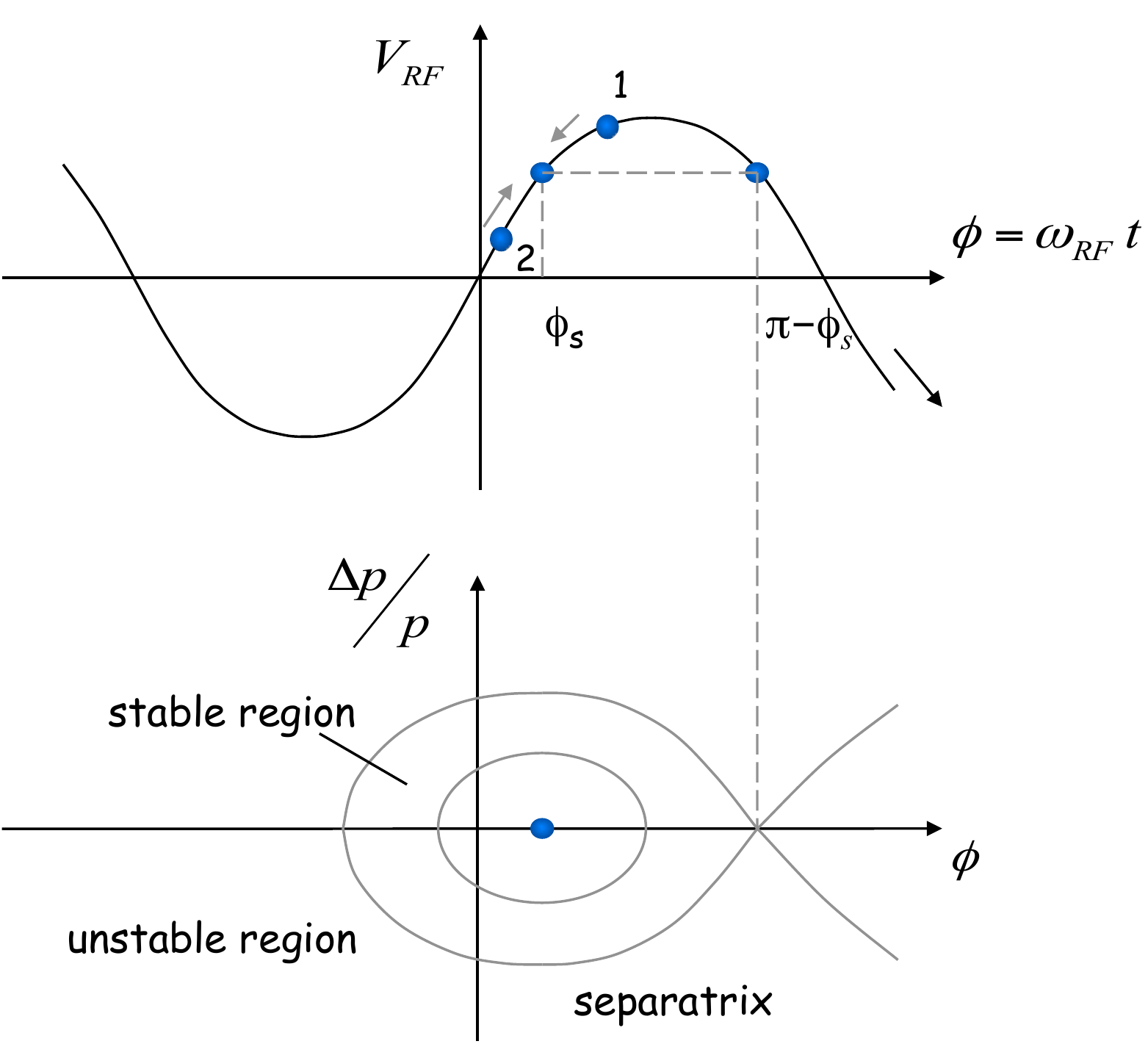}
 \caption{\label{fig:acc-bucket} RF voltage as a function of particle phase (top) and phase-space picture (bottom)}
\end{figure}

As we have seen before, the restoring force goes to zero when $\phi$ reaches $\pi - \phi_{\rm s}$ and becomes non-restoring beyond (both below and above transition); see Fig.~\ref{fig:acc-bucket}.
Hence, $\pi - \phi_{\rm s}$ is an extreme amplitude for a stable motion which has a closed trajectory in phase space. This phase-space trajectory separates the region of stable motion from the unstable region. It is called the {\it separatrix}. The area within this separatrix is called the {\it RF bucket}, which corresponds to the maximum acceptance in phase space for a stable motion.

Since we found an invariant of motion, we can write the equation for the separatrix by calculating it at the phase $\phi = \pi - \phi_{\rm s}$, where $\dot\phi = 0$:
\begin{equation}
\frac{\dot\phi^2}{2} - \frac{\Omega_{\rm s}^2}{\cos\phi_{\rm s}}  \left( \cos\phi + \phi \sin\phi_{\rm s} \right) = - \frac{\Omega_{\rm s}^2}{\cos\phi_{\rm s}}  \left( \cos(\pi -\phi_{\rm s}) + (\pi -\phi_{\rm s}) \sin\phi_{\rm s} \right).
\label{eq:separatrix}
\end{equation}

From this, we can calculate the second value $\phi_{\rm m}$ where the separatrix crosses the horizontal axis, which is the other extreme phase for stable motion:
 \begin{equation}
 \cos\phi_{\rm m} + \phi_{\rm m} \sin\phi_{\rm s}  = \cos(\pi -\phi_{\rm s}) + (\pi -\phi_{\rm s}) \sin\phi_{\rm s}.
\end{equation}

It can be seen from the equation of motion that $\dot\phi$ reaches an extreme when $\ddot\phi = 0$, corresponding to $\phi = \phi_{\rm s}$. Putting this value into equation~(\ref{eq:separatrix}) gives
\begin{equation}
\dot\phi^2_\mathrm{max} = 2 \,\Omega_{\rm s}^2 \left[ 2 + \left(2\phi_{\rm s}-\pi \right) \tan\phi_{\rm s} \right],
\end{equation}
which translates into an acceptance in energy
\begin{equation}
\left( \frac{\Delta E}{E_{\rm s}} \right)_\mathrm{max} = \pm \beta \sqrt{ - \frac{e \hat{V}}{\pi h \eta E_{\rm s}} G(\phi_{\rm s}) },
\end{equation}
where
\begin{equation}
G(\phi_{\rm s}) = 2 \cos\phi_{\rm s} + ( 2\phi_{\rm s}-\pi ) \sin\phi_{\rm s}.
\end{equation}

\begin{figure}[!htb]
 \centering\includegraphics[width=0.6\textwidth]{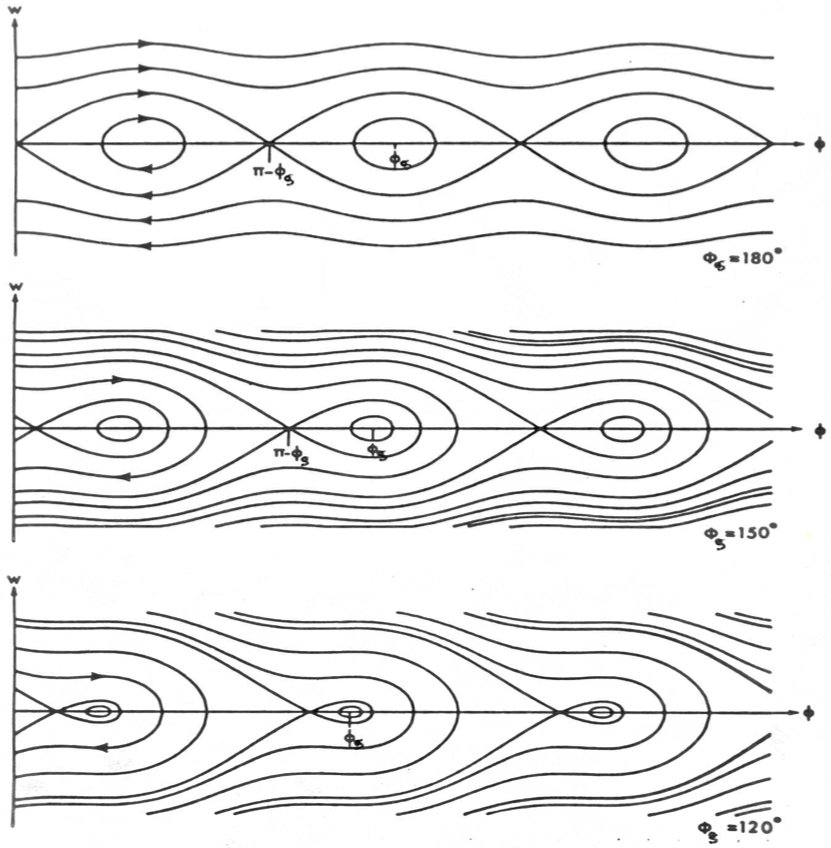}
 \caption{\label{fig:acceptance} Phase-space plots for different synchronous phase angles}
\end{figure}

This {\it RF acceptance} strongly depends on $\phi_{\rm s}$ and plays an important role for the capture at injection and the stored beam lifetime.
The maximum energy acceptance in the bucket depends on the square root of the available RF voltage $\hat{V}_{\rm RF}$.
The phase extension of the bucket is maximum for $\phi_{\rm s} = 180^\circ$.
As the synchronous phase gets closer to $90^\circ$ the bucket size becomes smaller, as illustrated in Fig.~\ref{fig:acceptance}.

\subsection{Stationary bucket}
In the case of the stationary bucket, we have no acceleration and $\sin\phi_{\rm s} = 0$, so that $\phi_{\rm s} = 0$ or $\pi$. The equation of the separatrix for $\phi_{\rm s} = \pi$ (above transition) simplifies to
\begin{equation}
\frac{\dot\phi^2}{2} + \Omega_{\rm s}^2 \cos\phi = \Omega_{\rm s}^2 \qquad \mathrm{or} \qquad \frac{\dot\phi^2}{2} = 2 \, \Omega_{\rm s}^2 \sin^2\frac{\phi}{2}.
\end{equation}

At this point, it is convenient to introduce a new variable $W$ to replace the phase derivative $\dot\phi$. As we see later, this new variable is canonical.
\begin{equation}
W = \frac{\Delta E}{\omega_{\rm rf}} = - \frac{p_{\rm s} R_{\rm s}}{h \eta \omega_{\rm rf}} \dot\phi.
\end{equation}
Introducing $\Omega_{\rm s}^2$ from equation~(\ref{eq:osc}) leads to the following equation for the separatrix:
\begin{equation}
W = \pm \frac{C}{\pi h c} \sqrt{ \frac{-e \hat{V} E_{\rm s}}{2 \pi h \eta}} \sin\frac{\phi}{2} = \pm W_{\rm bk} \sin\frac{\phi}{2} \qquad \mathrm{with} \qquad
W_{\rm bk} = \frac{C}{\pi h c} \sqrt{ \frac{-e \hat{V} E_{\rm s}}{2 \pi h \eta}}.
\end{equation}

Setting $\phi = \pi$ in the previous equation shows that $W_{\rm bk}$ is the maximum height of the bucket, which results in the maximum energy acceptance:
\begin{equation}
\Delta E_\mathrm{max} = \omega_{\rm rf} W_{\rm bk} = \beta_{\rm s} \sqrt{ 2 \frac{-e \hat{V}_{\rm RF} E_{\rm s}}{\pi h \eta}}.
\end{equation}

The bucket area is
\begin{equation}
A_{\rm bk} = 2 \int_0^{2\pi} W \mathrm{d}\phi.
\end{equation}
With $\int_0^{2\pi} \sin(\phi/2) \, \mathrm{d}\phi = 4$, one gets
\begin{equation}
A_{\rm bk} = 8 W_{\rm bk} = 8 \frac{C}{\pi h c} \sqrt{ \frac{-e \hat{V} E_{\rm s}}{2 \pi h \eta}}.
\end{equation}

\subsection{Bunch matching into the stationary bucket}
\begin{figure}[!ht]
 \centering\includegraphics[width=0.5\textwidth]{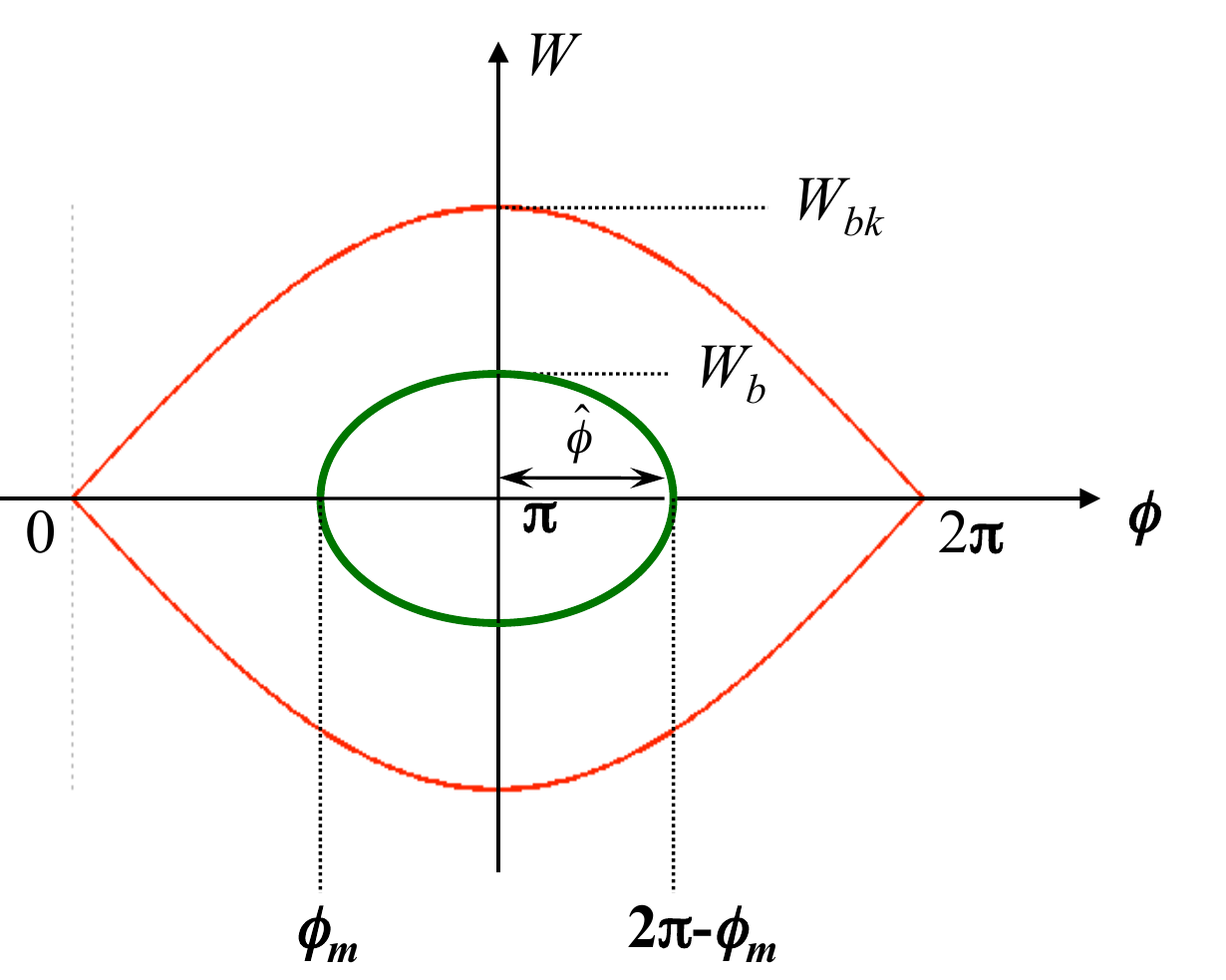}
 \caption{\label{fig:stationary-bucket} Phase-space plot for the separatrix of the stationary bucket and a trajectory inside}
\end{figure}

We can describe the motion of a particle inside the separatrix of the stationary bucket
starting from the invariant of motion from equation~(\ref{eq:invariant}) and setting $\phi_{\rm s} = \pi$:
\begin{equation}
\frac{\dot\phi^2}{2} + \Omega_{\rm s}^2 \,\cos\phi = I.
\end{equation}
The points $\phi_{\rm m}$ and $2\pi-\phi_{\rm m}$ where the trajectory crosses the horizontal axis are symmetric with respect to $\phi_{\rm s} = \pi$ (see Fig.~\ref{fig:stationary-bucket}).
We can calculate the invariant for $\phi = \phi_{\rm m}$ and get
\begin{equation}
\frac{\dot\phi^2}{2} + \Omega_{\rm s}^2 \,\cos\phi = \Omega_{\rm s}^2 \,\cos\phi_{\rm m},
\end{equation}
\begin{equation}
\dot\phi = \pm \Omega_{\rm s} \sqrt{2\, (\cos\phi_{\rm m} - \cos\phi)},
\end{equation}
\begin{equation}
W = \pm W_{\rm bk} \sqrt{\cos^2\frac{\phi_{\rm m}}{2} - \cos^2\frac{\phi}{2}} \qquad \left( \mathrm{using} \quad \cos\phi = 2 \cos^2\frac{\phi}{2} - 1 \right).
\end{equation}

Setting $\phi = \pi$ in the previous equation allows us to calculate the bunch height $W_{\rm b}$:
\begin{equation}
W_{\rm b} = W_{\rm bk} \,\cos\frac{\phi_{\rm m}}{2} = W_{\rm bk} \,\sin\frac{\hat\phi}{2}
\end{equation}
with $\hat\phi = \pi - \phi_{\rm m}$ being the maximum phase amplitude for an oscillation around the synchronous phase $\phi_{\rm s} = \pi$.

The corresponding maximum energy difference of a particle on this phase-space trajectory is
\begin{equation}
\left( \frac{\Delta E}{E_{\rm s}} \right)_{\!{\rm b}} = \left( \frac{\Delta E}{E_{\rm s}} \right)_{\rm RF} \cos\frac{\phi_{\rm m}}{2} = \left( \frac{\Delta E}{E_{\rm s}} \right)_{\rm RF} \sin\frac{\hat\phi}{2}.
\end{equation}

When a particle bunch is injected into a synchrotron, the bunch has a given bunch length and energy spread. The different particles will move along phase-space trajectories that correspond to their initial phase and energy. If the shape of the injected bunch in phase space matches the shape of a phase-space trajectory for the given RF parameters, the shape of the bunch in phase space will be maintained.

\begin{figure}[!b]
 \centering\includegraphics[width=\textwidth]{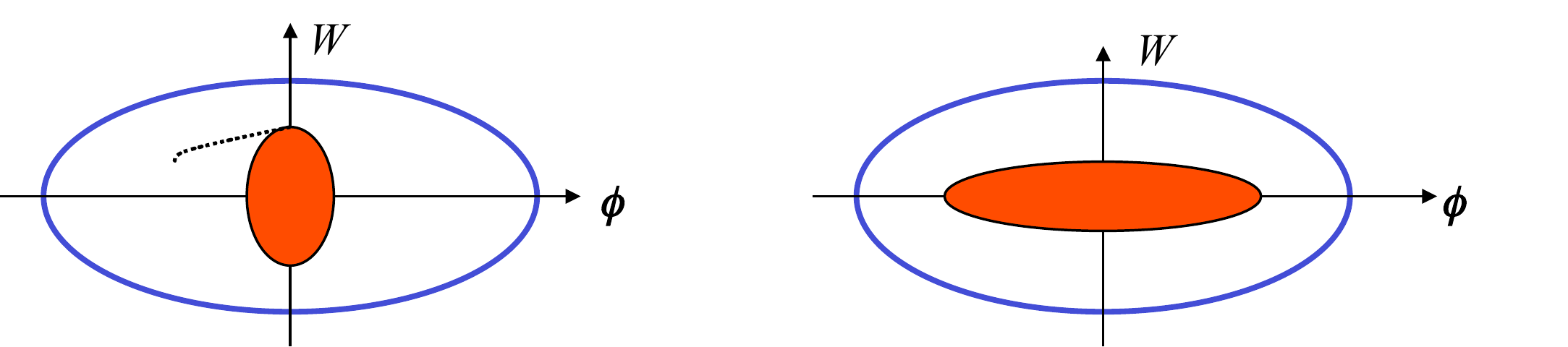}
 \caption{\label{fig:matching} Phase-space plots for a mismatched bunch 1/4 of a synchrotron period apart}
\end{figure}

If the shape does not match, it will vary during the synchrotron period. This is illustrated in Fig.~\ref{fig:matching} for a bunch that has a shorter bunch length and a higher energy spread compared to the phase-space trajectory. As the particles move on their individual trajectories, the bunch will be longer with a smaller energy spread after 1/4 of a synchrotron period, and will regain the initial shape after 1/2 period. This effect can be used to manipulate the shape of the bunch in phase space and trade off bunch length against energy spread (so-called {\it bunch rotation}).

\begin{figure}[!ht]
 \centering\includegraphics[width=0.9\textwidth]{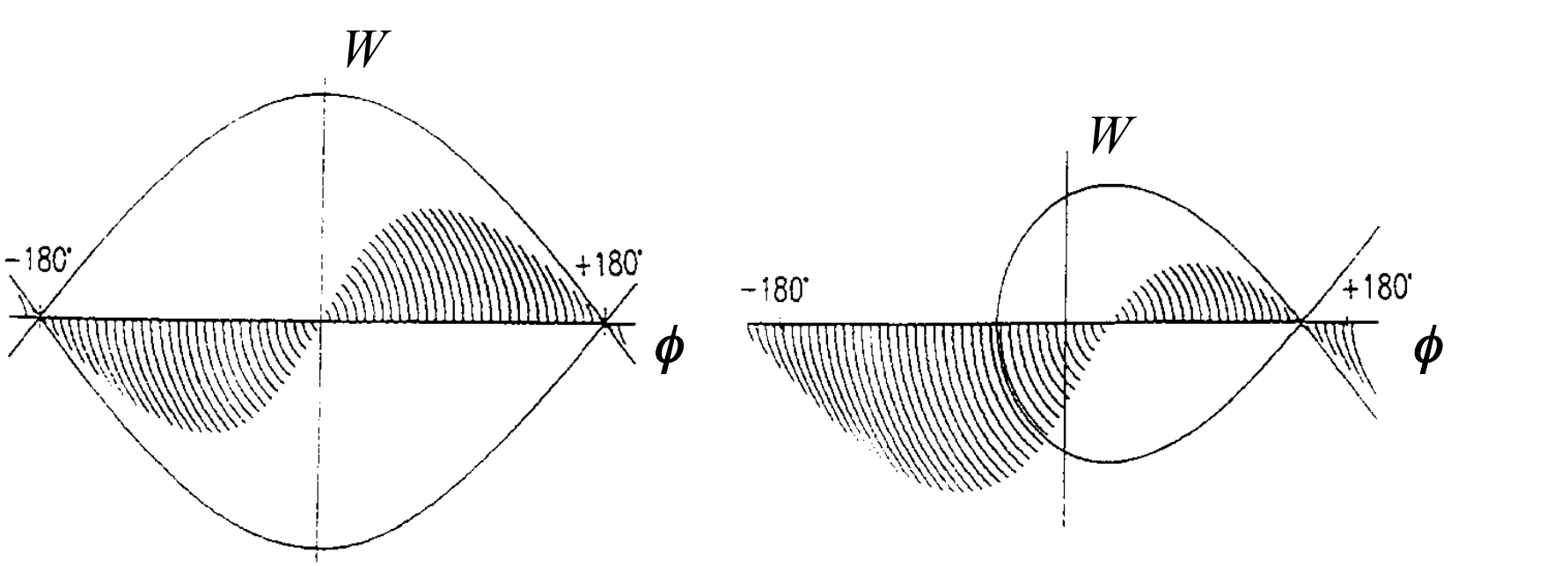}
 \caption{\label{fig:bucket-rotation} Phase-space trajectories for 1/8 of the synchrotron period for a stationary bucket (left) and an accelerating bucket (right) \cite{pirkl}.}
\end{figure}

Due to the non-linear restoring force, the synchrotron period depends on the oscillation amplitude, and particles with larger amplitudes have a longer synchrotron period, as shown in Fig.~\ref{fig:bucket-rotation}.
This will eventually lead to a {\it filamentation} of the bunch and an increase of the longitudinal emittance. The same will happen if a bunch is injected off-energy or with a phase error.

\subsection{Potential energy function and Hamiltonian}
The longitudinal motion is produced by a force that can be derived from a scalar potential $U$:
\begin{equation}
\der{^2\phi}{t^2} = F(\phi) = - \frac{\partial U}{\partial\phi},
\end{equation}
\begin{equation}
U(\phi) = - \int\limits_0^\phi F(\phi) \, \rd\phi = - \frac{\Omega^2_{\rm s}}{\cos\phi_{\rm s}} \left( \cos\phi + \phi\sin\phi_{\rm s} \right) - F_0.
\end{equation}
The sum of the potential energy and the kinetic energy is constant and by analogy represents the total energy of a non-dissipative system:
\begin{equation}
\frac{\dot\phi^2}{2} + U(\phi) = F_0.
\end{equation}

Since the total energy is conserved, we can describe the system as a Hamiltonian system. With the variable
\begin{equation}
W = \frac{\Delta E}{\omega_{\rm rf}} = \frac{R_{\rm s}}{h} \Delta p,
\end{equation}
the two first-order equations of the longitudinal motion become
\begin{eqnarray}
\der{\phi}{t} & = & - \frac{h \eta \omega_{\rm rf}}{p_{\rm s} R_{\rm s}} W \label{eq:dphi}, \\
\der{W}{t} & = & \frac{e \hat{V}}{2\pi h} \left( \sin\phi - \sin\phi_{\rm s} \right).
\end{eqnarray}

The two variables $\phi$ and $W$ are canonical since these equations of motion can be derived from a Hamiltonian $H(\phi, W, t)$:
\begin{equation}
\der{\phi}{t} = \pder{H}{W}, \qquad\der{W}{t} = - \pder{H}{\phi},
\end{equation}

\begin{equation}
H(\phi,W,t) = \frac{e \hat{V}}{2\pi h} \left[ \cos\phi - \cos\phi_{\rm s} + (\phi-\phi_{\rm s}) \sin\phi_{\rm s} \right] - \frac{1}{2} \frac{h \eta \omega_{\rm rf}}{R_{\rm s} p_{\rm s}} W^2.
\label{eq:ham}
\end{equation}

\subsection{Adiabatic damping}
So far it was assumed that the parameters $R_{\rm s}, p_{\rm s}, \omega_{\rm s}$, and $\hat{V}$ related to the longitudinal motion did not change appreciably over the time scale of a synchrotron period. Now consider that they vary slowly (adiabatically) with respect to the period of the longitudinal oscillation.
Then one needs to be more general in solving the second-order equation of motion~(\ref{eq:gen2}). There will be an additional term that is proportional to $\dot\phi$, indicating that there is a damping of the oscillation.

An elegant way to show the damping is possible with the help of the Boltzmann--Ehrenfest adiabatic theorem. This theorem states that for the canonically conjugate variables $p$ and $q$ of an oscillatory system with slowly changing parameters, the action integral $I$ taken over one period of oscillation remains constant:
\begin{equation}
I = \oint p \, \rd q = cte.
\end{equation}
We can apply this to the canonical variables $W$ and $\phi$ such that
\begin{equation}
I = \oint W \, \rd \phi = cte.
\end{equation}

For small-amplitude oscillations, the Hamiltonian from equation~(\ref{eq:ham}) becomes
\begin{equation}
H(\Delta\phi,W,t) \approxeq \frac{e \hat{V}}{4\pi h} \cos\phi_{\rm s} (\Delta\phi)^2 - \frac{1}{2} \frac{h \eta \omega_{\rm rf}}{R_{\rm s} p_{\rm s}} W^2
\end{equation}
with the solutions $W = \hat{W} \cos\Omega_{\rm s} t$ and $\Delta\phi = (\Delta\hat{\phi}) \sin\Omega_{\rm s} t$.
Using equation~(\ref{eq:dphi}), the action integral is given by
\begin{equation}
I = \oint W \der{\phi}{t} \, \rd t = - \frac{h \eta \omega_{\rm rf}}{R_{\rm s} p_{\rm s}} \oint W^2 \, \rd t.
\end{equation}
Taking the previous integral over one period leads, with
\begin{equation}
\oint W^2 \, \rd t = \pi \frac{\hat{W}^2}{\Omega_{\rm s}},
\end{equation}
to
\begin{equation}
I = - \frac{\pi h \eta \omega_{\rm rf}}{R_{\rm s} p_{\rm s}} \frac{\hat{W}^2}{\Omega_{\rm s}} = cte.
\end{equation}

From equation (\ref{eq:dphi}), one gets the relation between the amplitude $\hat{W}$ of the energy oscillation and the amplitude $\hat{\Delta\phi}$ of the phase oscillation:
\begin{equation}
\hat{W} = - \frac{R_{\rm s} p_{\rm s} \Omega_{\rm s}}{h \eta \omega_{\rm rf}} {\Delta\hat\phi}.
\end{equation}

So, one gets
\begin{equation}
{\Delta\hat\phi} \propto \left\{ \frac{\eta}{E_{\rm s} R_{\rm s}^2 \hat{V} \cos\phi_{\rm s}} \right\}^{1/4}. 
\end{equation}
Keeping all parameters constant except for the energy which is ramping, this means that the phase excursion ${\Delta\hat\phi}$ is decreasing with the one-fourth power of the energy. This relation also implies that the product $\hat{W} \cdot {\Delta\hat\phi}$ is invariant, so the phase-space area is preserved and Liouville's theorem still holds under adiabatically changing conditions. Only the shape of the phase-space ellipse is modified; the phase space is not damped.

The previous derivation does not include the damping by the emission of synchrotron radiation, as it occurs mainly for relativistic electrons and positrons. In this case, the longitudinal phase space will shrink due to the synchrotron radiation damping, and Liouville's theorem does not apply any more due to non-conservative forces.


\section*{Acknowledgment}

I wish to thank in particular Jo\"el Le Duff, who gave this lecture series in the past, and allowed me to use the material of his lectures, of which I profited a lot. Most of the figures and derivations of this lecture originate from him.

\section*{Bibliography}
J.~Le~Duff, in Proceedings of the CERN-CAS Accelerator School: 5th General Accelerator Physics Course, Jyvaskyla, Finland, 7--18 September 1992, CERN--1994-001 (CERN, Geneva, 1994), pp. 253-288, http://dx.doi.org/10.5170/CERN-1994-001.253.

\noindent J.~Le~Duff, in Proceedings of the CERN-CAS Accelerator School: 5th General Accelerator Physics Course, Jyvaskyla, Finland, 7--18 September 1992, CERN--1994-001 (CERN, Geneva, 1994), pp. 289-311, http://dx.doi.org/10.5170/CERN-1994-001.289.

\noindent H.~Wiedemann, {\sl Particle Accelerator Physics} (Springer, Berlin, 2007).

\noindent K.~Wille, {\sl The Physics of Particle Accelerators: An Introduction} (Oxford University Press, Oxford, 2000).

\noindent T.~Wangler, {\sl RF Linear Accelerators} (Wiley-VCH, Weinheim, 2008).

\end{document}